\documentclass[aps,preprint]{revtex4}%
\usepackage{amsfonts}
\usepackage{amsmath}
\usepackage{amssymb}
\usepackage{graphicx}%
\newtheorem{theorem}{Theorem}
\newtheorem{acknowledgement}[theorem]{Acknowledgement}

\begin{document}
\title[Gravitational quantum states of neutrons]{Gravitational quantum states of neutrons in a rough waveguide}
\author{A. E. Meyerovich}
\affiliation{Department of Physics, University of Rhode Island, Kingston, RI 02881-0817, USA}
\author{V. V. Nesvizhevsky }
\affiliation{ILL, Grenoble, France}

\begin{abstract}
A theory of gravitational quantum states of ultracold neutrons in waveguides
with absorbing/scattering walls is presented. The theory covers recent
experiments in which the ultracold neutrons were beamed between a mirror and a
rough scatterer/absorber. The analysis is based on a recently developed theory
of quantum transport along random rough walls which is modified in order to
include leaky (absorbing) interfaces and, more importantly, the low-amplitude
high-aperture roughness. The calculations are focused on a regime when the
direct transitions into the continuous spectrum above the absorption threshold
dominate the depletion of neutrons from the gravitational states and are more
efficient than the processes involving the intermediate states. The
theoretical results for the neutron count are sensitive to the correlation
radius (lateral size) of surface inhomogeneities and to the ratio of the
particle energy to the absorption threshold in a weak roughness limit. The
main impediment for observation of the higher gravitational states is the
"overhang" of the particle wave functions which can be overcome only by use
scatterers with strong roughness. In general, the strong roughness with high
amplitude is preferable if one wants just to detect the individual
gravitational states, while the strong roughness experiments with small
amplitude and high aperture are preferable for the quantitative analysis of
the data. We also discuss the ways to further improve the accuracy of
calculations and to optimize the experimental regime.

\end{abstract}
\volumeyear{year}
\volumenumber{number}
\issuenumber{number}
\eid{identifier}
\date[Date text]{date}
\received[Received text]{date}

\revised[Revised text]{date}

\accepted[Accepted text]{date}

\published[Published text]{date}

\startpage{1}
\endpage{ }
\maketitle
\tableofcontents

\section{Introduction}

One of the recent discoveries in neutron physics was the experimental
observation of quantization of motion of ultracold neutrons in gravitational
field \cite{neutron1}. By itself, the quantization of motion of particles by a
linear potential is a well-known textbook problem in quantum mechanics and has
already been observed for other ultracold particles (for example, for
spin-polarized atomic hydrogen in magnetic field with a linear gradient
\cite{freed1}). However, the Earth gravitational field is so weak and the
energies of the corresponding discrete quantum states for neutrons are so low
(of the scale of 1 \textrm{peV}!) that the observation of such states is
indeed a major experimental challenge. What is more, the measurements in this
energy range open the door, for instance, to experimental studies of the
fundamental short-range forces for neutrons near solid surfaces (see
\cite{pdg1,prot1} and references therein). This unique opportunity is inherent
to the high sensitivity of the neutron wave functions in the gravitationally
bound quantum states of neutrons to the existence of any new type of
short-range forces between a neutron and a surface.

The schematics of a typical experiment is the following. A collimated beam of
ultracold neutrons\ is sent between two, usually horizontal, solid plates at a
distance several $\mu\mathrm{m}$ from each other. The material of the plates
is such that the neutrons hitting the plates with the vertical velocity above,
approximately, \textrm{4 m/s}, get absorbed by the plates, while the particles
with a lower vertical component of velocity get reflected. One of the plates,
usually the upper one, has a rough surface while the other one is an almost
ideal mirror. The quality of the mirror is surprisingly good: the
single-reflection losses are in the range $10^{-4}\div10^{-5}$. As a result of
scattering by a rough wall, the direction of the neutron velocity changes and
the neutrons can acquire a sufficiently large vertical component of velocity
to get absorbed by the walls. Only the particles, which have extremely small
vertical velocity (low gravitational energy) do not reach the rough scatterer
(the upper plate) and, therefore, are not scattered and absorbed. These
particles get through the system and get counted by the exit counter. [In
practice, the bottom mirror is often made longer than the scatterer/absorber
and the entrance to the neutron counter is collimated with a long horizontal
narrow slit. With such configuration, the absorption is not the only source of
the neutron loss. The purpose of this slit is to avoid counting of neutrons
which exit at a large angle to the mirror surface and could otherwise get
counted, albeit with a small probability, either directly or after scattering
by spectrometer walls].

A quantum description of experiment differs from the classical one above in
several important aspects. The motion of neutrons in the gravitational field
between the plates is quantized. The lower the energy, the smaller the
(vertical) size of the state. The particles in the lowest gravitational states
do not reach the scatterer (or, more precisely, reach it with an exponentially
low probability), do not scatter, and, therefore, do not get absorbed. With
decreasing distance between the plates, the lower and lower states will reach
the scatterer and get absorbed. As a result, the neutron count in the case of
spatial quantization of neutrons, should, in ideal circumstances, be a
step-wise function of the clearance between the plates. The positions of the
steps should then provide the calibration for ultralow energy measurements. In
experiment, unfortunately, such a clear-cut step-wise picture does not emerge:
though the curves, without doubts, demonstrate the quantization of the
gravitational states, the higher gravitational states have, so far, eluded identification.

One of the main purposes of this paper is to develop a general theory that
would cover the experiment, explain the difficulties associated with
identification of the higher states, and suggest the ways to optimize the
experimental regime. Many aspects of the problem have been discussed very
recently in Ref. \cite{voronin1} which analyzed potential mechanisms of
neutron losses in experiment \cite{neutron1}, including the surface roughness
as the principal mechanism of non-specularity of reflection which is
ultimately responsible for the neutron depletion. This paper demonstrated that
the losses of neutrons from the lowest gravitational states are determined by
the neutron tunneling through the gravitational barrier between the classical
turning point and the position of the scatterer/absorber.

However, the theory \cite{voronin1} did not provide an explicit quantitative
link between the rate of neutron loss and parameters of the roughness of the
neutron waveguide. Since the non-specularity of reflection by the rough
surfaces is an obvious principal source of the neutron depletion in experiment
\cite{neutron1}, any accurate quantitative description of the phenomenon
should express the neutron count directly via the parameters of the random
surface roughness. Our results below not only provide such an explicit link,
but also suggest what kind of roughness is the most advantageous for experiment.

In order to extract quantitative information from the experiment and,
eventually, to study the short-range forces, one needs an accurate theoretical
description of quantum diffusion of neutrons along rough walls. Recently we
developed a theory of quantum transport in quasi-$2D$ systems with random
rough boundaries (see Refs. \cite{arm2,pon1} and references therein). In this
paper we apply this formalism to quantized neutrons in the gravitational
field. The practical purpose is twofold: to develop an accurate quantitative
theory of neutron transport in the presence of random surfaces, which can be
used for a quantitative analysis of experimental data, and to determine the
parameters of surface roughness for the scatterer surface that would help to
optimize future experiments. As for the purely theoretical benefits, we are
able to extend our general theory to include strong, high-aperture roughness
and to develop a theoretical description for systems with absorbing, "leaky" walls.

Roughness of the walls affects the quantized, in this case, gravitational,
states of particles inside the quantum wells by shifting and broadening the
energy levels. The stronger the roughness, the larger are these shifting and
broadening. In this paper we, effectively, calculate the line broadening which
is determined by the roughness-driven transition between the states. Our
perturbative approach works as long as the resulting uncertainty in the level
positions is smaller than the distance between the levels. The same condition
is necessary for the quantitative interpretation of experiment: though strong
roughness can make the effect of the gravitational quantization of states more
pronounced, the uncertainty in the level positions can render the results
unusable for the calibration of the short-range forces.

The strength of roughness is characterized by the ratio of its amplitude
$\ell$ to the size of the well $L$ and by the aperture of roughness $\ell/R$,
which is determined by the ratio of the amplitude of roughness $\ell$ to its
typical lateral scale $R$ (the correlation radius of surface inhomogeneities).
Usually, both the line shifts and broadening are small as long as $\ell
/L,\ell/R\ll1$ (see, for example, Ref. \cite{arm2} and references therein).

In this particular case there could be two exceptions. First, the highest
energy levels, which are close to the absorption threshold, could become
relatively broad even for a moderate roughness. This is not very important for
the results below since in this paper we deal mostly with direct
roughness-driven transitions from the gravitational states into the continuous
spectrum above the threshold and can disregard the transitions via the
intermediate states. Second, and more important, the effect of surface
roughness on the lowest gravitational states is anomalously small because for
such states the particle wave functions on the rough walls are exponentially
small in the distance between the classical turning point and the wall.
Therefore, as long as we can disregard the absorption processes which involve
the intermediate, higher levels, the linewidths of the lowest gravitational
states remain small and the calculation of the absorption rate can even be
extended to stronger, high-aperture roughness. We plan to address the neutron
absorption that involves the transitions via the intermediate states later.

When interpreting the experimental results, the main uncertainty is introduced
by the lack of accurate data on parameters of surface roughness, especially on
the type of correlations and the correlation length of surface
inhomogeneities. Simple measurements of the average amplitude and the lateral
size of roughness are insufficient when one needs to know the shape of the
correlation function of surface inhomogeneities which can be extracted, almost
exclusively, from diffraction measurements \cite{q2}. Another major sort of
uncertainty is a relatively poor information on the energy/velocity
distribution of neutrons in the beam. While the dispersion in horizontal
component of velocity can be measured very accurately, the precise
measurements of the vertical dispersion are much more problematic. Therefore,
one of the goals of this paper is to identify the range of parameters in which
this lack of precise input data affects the results as little as possible.
Surprisingly, this is possible.

The paper has the following structure. In the next Section we describe the
neutron wave functions and eigenstates, introduce proper dimensionless
variables, and present a set of transport equations for quantized neutrons
near random rough wall(s). In Section III we present the roughness-driven
scattering probabilities in the case of weak roughness. In Section IV we
explore the ways to generalize our theory to stronger roughness with some
success in the case of high-aperture low-amplitude roughness. The final
results for absorption times for neutrons on the leaky walls and for the
neutron count for direct and inverse geometries are given in Section V.
Section VI contains conclusions and discussion of how to improve our theory
and optimize the experiment.

\section{Main equations}

\subsection{Potential well, notations, and dimensionless variables}

In the bulk of the paper we will disregard the ultrashort-range potential for
neutrons near solid surfaces and consider surfaces of solid plates as
potential barriers for neutrons of the height $U_{c}$ (the absorption
barrier). We will discuss the short-range forces near the walls towards the
end of the paper. If one forgets for a moment about the roughness of the
scatterer/absorber (the upper plate), the potential energy for neutrons in the
vertical direction, $U\left(  z\right)  ,$ is $U\left(  z\leq0\right)
=U_{c}+mgz$ (mirror at the bottom), $U\left(  z\right)  =mgz$ at $0<z<H$ (the
gravitational field between the plates; the scatterer/absorber is at the
distance $H$ from the bottom mirror), and $U\left(  z\right)  =mgz+U_{c}\equiv
U_{H}+mg\left(  z-H\right)  $ (the absorption barrier) at $z\geq H$.
Schematically, this potential is presented in Figure 1 by the dashed lines. An
alternative geometry is discussed in Section VE.%

\begin{figure}
[ptb]
\begin{center}
\includegraphics[
natheight=4.076700in,
natwidth=5.280500in,
height=3.5388in,
width=4.5766in
]%
{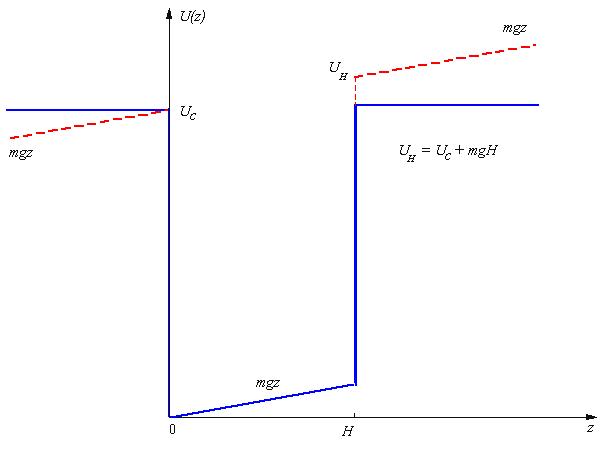}%
\caption{Schematic drawing of the potential well in vertical direction $z$.
Dashed lines - the "real" potential; solid lines - the potential used in the
paper. Since $mgH/U_{c}\sim10^{-5}$ and we are particularly interested in the
lowest energy levels, the corrections are negligible [with an accurately
scaled potential, the difference between the dashed and solid levels cannot be
seen].}%
\end{center}
\end{figure}

Numerically, $U_{c}\sim1.\,\allowbreak34\times10^{-26}$ \textrm{J} (the
kinetic barrier for absorption is $4$ \textrm{m/s}). The characteristic
gravitational energies inside the well, on the other hand, are by five orders
of magnitude lower, $mgH\sim1.\,\allowbreak6\times10^{-31}$ \textrm{J
}(for\textrm{ }$H\sim10$ $\mu\mathrm{m}$). This means that the plot in Figure
1 is out of scale. Since we are interested almost exclusively in the
gravitational states near the bottom of the well, we can, without any
noticeable loss of accuracy, ignore the gravitational field near the top of
the well and replace the exact potential (the dashed lines) by the one given
in Figure 1 by solid lines.

The energy spectrum of neutrons in this potential well consists of discrete
energy states $\epsilon_{n}$ below $U_{c}$ and continuum above. [Note, that
the short-lived discrete states near the threshold could be also treated as
continuum. In the well walls are not ideal, but rough, all the energy levels,
especially the higher ones, broaden. With an increase in roughness amplitude,
lower and lower levels will acquire a noticeable width. This is, in some
sense, equivalent to the lowering of the absorption threshold with increasing
roughness]. We will call the discrete states close to the bottom of the well
the "gravitational states" and higher states in the main part of the well -
the "square well states". Though this terminological shortcut is not
well-defined, it is convenient for discussing the physics and the results
because of a very large ratio $U_{c}/mgH\sim10^{5}$; in computations, unless
stated otherwise, all the states will, of course, be treated accurately.

It is convenient to measure the distances $z$ and the energies $\epsilon_{n}$
in units of $l_{0}$ and $e_{0}$, $\lambda_{n}=\epsilon_{n}/e_{0}$ and
$s=z/l_{0}$, where $l_{0}=\hbar^{2/3}\left(  2m^{2}g\right)  ^{-1/3}\sim5.871$
$\mu\mathrm{m}$ and $e_{0}=mgl_{0}\sim0.602$ \textrm{peV}$\sim9.6366\times
10^{-32}$ $\mathrm{J}$ are the size and the gravitational energy of a neutron
in the lowest quantum state in the \emph{infinite} gravitational trap. In
these units, the overall kinetic energy of particles in the beam $E$ and the
barrier height $U_{c}$ are $1.4\times10^{5}<e=E/e_{0}<$ $8.7\times10^{5},$
$u_{c}=U_{c}/e_{0}\sim1.4\times10^{5}$. A very important parameter is the
ratio $\chi=u_{c}/e$. In experiment,
\begin{equation}
0.15\lesssim\chi\equiv u_{c}/e<1. \label{k1}%
\end{equation}
The value of this parameter shows to what extent the kinetic energy of a
neutron exceeds the barrier and determines how easy is it for a neutron to get
absorbed by a plate when the direction of velocity is rotated as a result of
scattering by the rough scatterer/absorber. As we will see, the results are
very sensitive to the exact value of $\chi$ even within the limited range
$\left(  \ref{k1}\right)  $ because both values of $u_{c}$ and $e$ are very large.

It is also convenient to introduce the dimensionless velocities (momenta) in
the beam direction along the well ($x$ direction) $\beta_{j}$, $v_{j}%
=\beta_{j}v_{0},\ v_{0}=\sqrt{2gl_{0}}\sim1.073\times10^{-2}\ \mathrm{m/s},$
$\beta_{j}=\sqrt{e-\lambda_{j}}\equiv p_{j}l_{0}\approx\sqrt{e}\left(
1-\frac{1}{2}\lambda_{j}/e\right)  =\sqrt{e-\lambda_{j}}$ (for lower levels,
$\lambda_{j}\ll e$ and $\beta_{j}\simeq\sqrt{e}\left(  1-\frac{1}{2}%
\lambda_{j}/e\right)  $). The range of kinetic energies in experiment is
$373<\sqrt{e}<932.$

The roughness of the surface of the scatterer/absorber (the upper plate) is
characterized by the correlation function of surface inhomogeneities which is
determined by the average lateral size (correlation radius) $R$ and the
average amplitude (height) $\ell$ of the inhomogeneities (for precise
definitions see Section III). In dimensionless variables, $r=R/l_{0}$,
$\eta=\ell/l_{0}.$ It is reasonable to expect that $\eta\lesssim0.1$ [so far,
in experiments \cite{neutron1} $0.03\lesssim\eta\lesssim0.1$; roughness with
higher amplitude than 0.1 would be too strong for a quantitative theory]. We
will discuss how to introduce the surface roughness into the equations later on.

\subsection{Wave functions and energies of the bound states}

At $E<U_{c}$ (or $\lambda_{n}<u_{c}\equiv U_{c}/e_{0}$) the wave function
$\psi_{n}\left(  s\right)  $, $s=z/l_{0}$, inside the well, $\ 0\leq s\leq
h=H/l_{0}$, is
\begin{equation}
\psi_{n}\left(  s\right)  =A_{n}\cdot\left[  \mathrm{Ai}\left(  s-\lambda
_{n}\right)  -S_{n}\mathrm{Bi}\left(  s-\lambda_{n}\right)  \right]  .
\label{n1}%
\end{equation}
Outside the well,%
\begin{align}
\psi_{n}\left(  s\right)   &  =B_{n}\cdot\exp\left[  -k_{n}\left(  s-h\right)
\right]  ,\ s>h,\label{n3}\\
\ \psi_{n}\left(  s\right)   &  =C_{n}\cdot\exp\left[  k_{n}s\right]
,\ s<0,\ k_{n}=\sqrt{u_{c}-\lambda_{n}}.\nonumber
\end{align}
The values of parameters $S_{n}$ and the energy eigenvalues $\lambda_{n}$ are
determined by the continuity conditions at $s=0$ and $s=h$:
\begin{align}
C_{n}  &  =A_{n}\cdot\left[  \mathrm{Ai}\left(  -\lambda_{n}\right)
-S_{n}\mathrm{Bi}\left(  -\lambda_{n}\right)  \right]  ,\ B_{n}=A_{n}%
\cdot\left[  \mathrm{Ai}\left(  h-\lambda_{n}\right)  -S_{n}\mathrm{Bi}\left(
h-\lambda_{n}\right)  \right]  ,\label{nn3}\\
C_{n}k_{n}  &  =A_{n}\cdot\left[  \mathrm{Ai}^{\prime}\left(  -\lambda
_{n}\right)  -S_{n}\mathrm{Bi}^{\prime}\left(  -\lambda_{n}\right)  \right]
,\ B_{n}k_{n}=-A_{n}\cdot\left[  \mathrm{Ai}^{\prime}\left(  h-\lambda
_{n}\right)  -S_{n}\mathrm{Bi}^{\prime}\left(  h-\lambda_{n}\right)  \right]
,\nonumber
\end{align}
which are equivalent to two equations%
\begin{align}
1/k_{n}  &  =\left[  \mathrm{Ai}\left(  -\lambda_{n}\right)  -S_{n}%
\mathrm{Bi}\left(  -\lambda_{n}\right)  \right]  /\left[  \mathrm{Ai}^{\prime
}\left(  -\lambda_{n}\right)  -S_{n}\mathrm{Bi}^{\prime}\left(  -\lambda
_{n}\right)  \right] \label{nnn4}\\
&  =-\left[  \mathrm{Ai}\left(  h-\lambda_{n}\right)  -S_{n}\mathrm{Bi}\left(
h-\lambda_{n}\right)  \right]  /\left[  \mathrm{Ai}^{\prime}\left(
h-\lambda_{n}\right)  -S_{n}\mathrm{Bi}^{\prime}\left(  h-\lambda_{n}\right)
\right]  .\nonumber
\end{align}
Computation of the neutron count would also require the normalization
coefficients $A_{n},$%
\begin{align}
A_{n}^{-2}  &  =\frac{1}{l_{0}}%
{\displaystyle\int\nolimits_{0}^{h}}
\left[  \mathrm{Ai}\left(  s-\lambda_{n}\right)  -S_{n}\mathrm{Bi}\left(
s-\lambda_{n}\right)  \right]  ^{2}ds\label{nnn55}\\
&  +\frac{1}{2l_{0}k_{n}}\left[  \left[  \mathrm{Ai}\left(  -\lambda
_{n}\right)  -S_{n}\mathrm{Bi}\left(  -\lambda_{n}\right)  \right]
^{2}+\left[  \mathrm{Ai}\left(  h-\lambda_{n}\right)  -S_{n}\mathrm{Bi}\left(
h-\lambda_{n}\right)  \right]  ^{2}\right]  ,\nonumber
\end{align}
and $B_{n},C_{n}$, Eq. $\left(  \ref{nn3}\right)  $. Some information on
discrete states can be also found in Ref. \cite{voronin1}.

\subsection{Approximate description of the bound states}

For what follows, we will not always need the exact eigenvalues and the wave
functions. First, let us note that the threshold energy $U_{c}$ is much higher
(by about five orders of magnitude!) than the gravitational energy $mgH$ in a
typical experiment, $u_{c}\gg h$. One can also separate the states into two
groups: the lowest states for which the presence of the gravitational field
near the bottom of the well is important (the "gravitational states") and the
higher states for which the fact that the gravity determines the shape of the
bottom of the well is mostly irrelevant (the "square well" states).

For the gravitational states in an infinite well, $u_{c}\rightarrow\infty,$
the wave functions $\left(  \ref{n1}\right)  $\ should become equal to zero on
the walls $s=0;h$, and equations for the eigenvalues Eq. $\left(
\ref{nnn4}\right)  $ should be replaced by the following equation for the
energy spectrum $\overline{\lambda}_{n}\left(  h\right)  $ [here and below we
identify physical parameters, which are calculated for a well with infinite
walls, by a bar over the corresponding symbols]:%
\begin{equation}
\overline{S}_{n}=\mathrm{Ai}\left(  -\overline{\lambda}_{n}\right)
/\mathrm{Bi}\left(  -\overline{\lambda}_{n}\right)  ,\ \mathrm{Ai}\left(
h-\overline{\lambda}_{n}\right)  -\overline{S}_{n}\mathrm{Bi}\left(
h-\overline{\lambda}_{n}\right)  =0. \label{nn6}%
\end{equation}
The normalizing coefficients $A_{n}$ are now defined as%
\begin{align}
\overline{A}_{n}^{2}  &  =\frac{1}{l_{0}}a_{n},\ a_{n}=\left\{
{\displaystyle\int\nolimits_{0}^{h}}
\left[  \mathrm{Ai}\left(  s-\overline{\lambda}_{n}\right)  -\overline{S}%
_{n}\mathrm{Bi}\left(  s-\overline{\lambda}_{n}\right)  \right]
^{2}ds\right\}  ^{-1}\label{nn7}\\
&  =\left\{  \left(  \mathrm{Ai}^{\prime}\left(  -\overline{\lambda}%
_{n}\right)  -\overline{S}_{n}\mathrm{Bi}^{\prime}\left(  -\overline{\lambda
}_{n}\right)  \right)  ^{2}-\left(  \mathrm{Ai}^{\prime}\left(  h-\overline
{\lambda}_{n}\right)  -\overline{S}_{n}\mathrm{Bi}^{\prime}\left(
h-\overline{\lambda}_{n}\right)  \right)  ^{2}\right\}  ^{-1}.\nonumber
\end{align}
If for the lowest gravitational states the size of the state $\lambda_{n}$ is
much smaller than the spacing between the walls, $\lambda_{n}\ll h$, the
equation for the energy spectrum is even simpler,%
\begin{equation}
\mathrm{Ai}\left(  -\widetilde{\lambda}_{n}\right)  =0. \label{nn8}%
\end{equation}
with the semiclassical eigenvalues
\[
\widetilde{\lambda}_{n}^{sc}=\left(  \frac{3\pi}{4}\left(  2n-\frac{1}%
{2}\right)  \right)  ^{2/3}%
\]
and the normalizing coefficients%
\begin{equation}
\widetilde{A}_{n}^{2}=\frac{1}{l_{0}}\widetilde{a}_{n},\ \widetilde{a}%
_{n}=\left[
{\displaystyle\int\nolimits_{0}^{\infty}}
\mathrm{Ai}\left(  s-\widetilde{\lambda}_{n}\right)  ^{2}ds\right]
^{-1}=\frac{1}{l_{0}}\mathrm{Ai}^{\prime-2}\left(  -\widetilde{\lambda}%
_{n}\right)  . \label{nn10}%
\end{equation}

Since $u_{c}$ is very large but finite, $u_{c}\sim10^{5}$,\ the eigenvalues
$\lambda$ differ from their values $\overline{\lambda}$ $\left(
\ref{nn6}\right)  $ at $u_{c}\rightarrow\infty$, $\lambda=\overline{\lambda
}+\delta\lambda,\ S=\overline{S}+\delta S$ $.$ The corresponding corrections
can be calculated by expanding Eqs. $\left(  \ref{nnn4}\right)  $ in
$1/\sqrt{u_{c}-\overline{\lambda}}$:%
\begin{align}
-\frac{1}{\sqrt{u_{c}-\overline{\lambda}}}  &  =\delta\lambda+\delta
S\frac{\mathrm{Bi}\left(  -\overline{\lambda}\right)  }{\mathrm{Ai}^{\prime
}\left(  -\overline{\lambda}\right)  -\overline{S}\mathrm{Bi}^{\prime}\left(
-\overline{\lambda}\right)  },\label{nnn10}\\
\frac{1}{\sqrt{u_{c}-\overline{\lambda}}}  &  =\delta\lambda+\delta
S\frac{\mathrm{Bi}\left(  h-\overline{\lambda}\right)  }{\mathrm{Ai}^{\prime
}\left(  h-\overline{\lambda}\right)  -\overline{S}\mathrm{Bi}^{\prime}\left(
h-\overline{\lambda}\right)  }.\nonumber
\end{align}
Then the normalizing coefficients $B,C$ ($\overline{B}=\overline{C}=0$)$,$
become%
\begin{align}
B  &  =\psi\left(  H\right)  =-\overline{A}\left[  \mathrm{Ai}^{\prime}\left(
h-\overline{\lambda}\right)  -\overline{S}\mathrm{Bi}^{\prime}\left(
h-\overline{\lambda}\right)  \right]  /\sqrt{u_{c}-\overline{\lambda}%
},\label{nn14}\\
C  &  =\psi\left(  0\right)  =\overline{A}\left[  \mathrm{Ai}^{\prime}\left(
-\overline{\lambda}\right)  -\overline{S}\mathrm{Bi}^{\prime}\left(
-\overline{\lambda}\right)  \right]  /\sqrt{u_{c}-\overline{\lambda}%
}.\nonumber
\end{align}

For the square well states well above the bottom, but still below $U_{c},$ one
can neglect the gravitational field near the bottom and use the standard
equation for the eigenvalues $\lambda_{n},$%
\begin{align}
\tan\left(  \sqrt{\lambda}h\right)   &  =-\frac{\sqrt{\lambda}+kS}%
{k-\sqrt{\lambda}S},\label{nnn14}\\
k  &  \equiv\sqrt{u_{c}-\lambda},\ S\equiv\sqrt{\lambda/\left(  u_{c}%
-\lambda\right)  }. \label{nnn16}%
\end{align}
For these symmetric states the values of the wave functions on the walls are%
\begin{align}
\psi_{n}^{2}\left(  0\right)   &  =\psi_{n}^{2}\left(  H\right)  =A^{2}%
S^{2},\label{nnn17}\\
A^{-2}  &  =\frac{l_{0}}{k}S^{2}+\frac{l_{0}}{4\sqrt{\lambda}}\left[  2\left(
\sqrt{\lambda}h+S+\sqrt{\lambda}hS\right)  ^{2}-2S\cos\left(  2\sqrt{\lambda
}h\right)  -\left(  1-S^{2}\right)  \sin\left(  2\sqrt{\lambda}h\right)
\right]  .\nonumber
\end{align}
In the case of very deep levels $\lambda/u_{c}$, Eqs. $\left(  \ref{nnn14}%
\right)  -\left(  \ref{nnn17}\right)  $ can be simplified as%
\begin{align}
\lambda &  =\overline{\lambda}\left(  1-\frac{4}{h\sqrt{u_{c}}}\right)
,\ S=\sqrt{\frac{\overline{\lambda}}{u_{c}}},\label{nnn18}\\
\psi_{n}^{2}\left(  0\right)   &  =\psi_{n}^{2}\left(  H\right)  =\frac
{2}{l_{0}h}\frac{\overline{\lambda}}{u_{c}}, \label{nnn19}%
\end{align}
where $\overline{\lambda}$ and $\overline{A}$\ are the eigenvalues and the
normalizing coefficient for the infinitely deep well,
\begin{equation}
\overline{\lambda}_{n}=\pi^{2}n^{2}/h^{2},\ \overline{A}_{n}^{2}=1/\left(
2l_{0}h\right)  . \label{nn11}%
\end{equation}

Approximately, the total number of discrete levels is%
\begin{equation}
\mathcal{N}\sim\frac{h}{\pi}\sqrt{u_{c}}\approx120h. \label{nnn11}%
\end{equation}

\subsection{Time evolution of the neutron population}

Recently we developed a versatile formalism for description of $2D$ diffusion
of quantized particles along random rough walls \cite{arm2}. The key is to
include the information on the wall roughness not into the boundary
conditions, but into the roughness-driven transition probabilities between the
quantum states in the transport equation. The application of this formalism to
the neutron problem at hand requires several important modifications: we
should include neutron absorption by the "leaky" walls and reconsider the
time/space properties of the transport equation.

The possibility of neutron absorption by the walls can be taken into account
by the calculation of the probability of transitions into the continuous
spectrum without the counterbalancing terms in the transport equation. The
second anomaly of this problem is that we are calculating not a stationary
diffusion flow but the time dynamics of the neutron count. As a result, we
have to solve a time-dependent problem with an initial condition in the form
of initial distribution of neutrons over the vertical velocities, \emph{i.e.,}
over the gravitational quantum states.

The main modification deals with the dimensionality of the problem. In
\cite{arm2} we considered the scattering-driven $2D$ diffusion in two
equivalent directions parallel to the walls while the third dimension (the
finite motion perpendicular to the plates) gave rise to a set of discrete
quantum states. In our neutron problem there is a preferred direction along
the plates (the beam direction $x$); the neutron count is being done only in
this direction. Therefore, we should exclude the sideway motion and
reformulate the transport problem as a $2D$ problem with vertical motion
(coordinate $z$) and the direction of the beam ($x$-direction with the
corresponding momentum $p$). In other words, we have to modify the transport
equation of Refs. \cite{arm2,pon1} (equations for the particle balance) for
quasi-$2D$ particles (particles with continuous spectrum in $x,y$ directions
and discrete quantized gravitational states in $z$ direction) in order to
eliminate the momentum in $q$ direction $y$.

According to \cite{arm2}, the distribution functions $n_{j}\left(  q,p\right)
$ in each quantized state $j$ obey the following Boltzmann equations:
\begin{equation}
\frac{dn_{j}}{dt}=2\pi\sum_{,j^{\prime}}\int W_{jj^{\prime}}\left(
q,p;q^{\prime},p^{\prime}\right)  \left[  n_{j^{\prime}}-n_{j}\right]
\delta\left(  \epsilon_{jqp}-\epsilon_{j^{\prime}q^{\prime}p^{\prime}}\right)
\frac{dq^{\prime}dp^{\prime}}{\left(  2\pi\right)  ^{2}}. \label{e6}%
\end{equation}
where the $2D$ vector $\mathbf{q}$ in the plane of the mirror has $y$ and $x$
components $\mathbf{q=}\left(  q,p\right)  $. Since we disregard the change of
momentum in $y$-direction,%
\begin{equation}
W_{jj^{\prime}}\left(  \mathbf{q,q}^{\prime}\right)  =\delta\left(
q-q^{\prime}\right)  W_{jj^{\prime}}\left(  p\mathbf{,}p^{\prime}\right)
,\ n_{j}\left(  q,p\right)  =\frac{2\pi}{L_{y}}\delta\left(  q\right)
n_{j}\left(  p\right)  , \label{ee6}%
\end{equation}
and the integration over $dq^{\prime}$ in Eq.$\left(  \ref{e6}\right)  $
yields
\begin{equation}
\frac{dn_{j}\left(  p\right)  }{dt}=2\pi\sum_{,j^{\prime}}\int W_{jj^{\prime}%
}\left(  p\mathbf{,}p^{\prime}\right)  \left[  n_{j^{\prime}}\left(
p^{\prime}\right)  -n_{j}\left(  p\right)  \right]  \delta\left(  p^{\prime
2}/2m-p^{2}/2m+\epsilon_{j^{\prime}}-\epsilon_{j}\right)  \frac{dp^{\prime}%
}{2\pi}. \label{e7}%
\end{equation}
We are dealing with particles with a fixed energy $E$, \emph{i.e.,} with
particles in gravitational quantum states $\epsilon_{j}$ with distinct values
of the lateral momentum $p_{j}=\sqrt{2m\left(  E-\epsilon_{j}\right)  }$,
\begin{equation}
n_{j}\left(  p\right)  =\frac{2\pi}{L_{x}}\delta\left(  p-p_{j}\right)  N_{j},
\label{e8}%
\end{equation}
where $N_{j}$ is the number of particles in state $j$ per unit length of the
beam. Finally, the equations for the particle balance $\left(  \ref{e7}%
\right)  $ reduce to
\begin{equation}
\frac{dN_{j}}{dt}=\sum_{,j^{\prime}}W_{jj^{\prime}}\left(  p_{j}%
\mathbf{,}p_{j^{\prime}}\right)  \left[  N_{j^{\prime}}/v_{j^{\prime}}%
-N_{j}/v_{j}\right]  . \label{e9}%
\end{equation}
To this equation we should add the terms responsible for the neutron
absorption (transitions into the continuous spectrum):%
\begin{equation}
\partial N_{j}/\partial t=\sum_{\lambda_{j^{\prime}}\leq u_{c}}W_{jj^{\prime}%
}\left(  N_{j^{\prime}}/v_{j^{\prime}}-N_{j}/v_{j}\right)  -N_{j}/v_{j}%
{\displaystyle\int\nolimits_{\lambda^{\prime}>u_{c}}}
W_{\lambda^{\prime}j}d\lambda^{\prime}. \label{n6}%
\end{equation}

Note, that the separation of the energy spectrum into discrete lines and
continuum, which is exact for an ideal well in Figure 1, is unambiguous only
for weak roughness. In general, roughness leads to shifting and broadening of
all the lines. The transition probabilities and, therefore, the broadening are
expected to be higher for the higher levels (see the next Section). With
increasing amplitude of roughness, the higher levels can become broad enough
to be considered continual, which, in turn, is equivalent to the lowering of
the absorption threshold. In the case of the high-amplitude roughness this
effect could even become dominant. However, in the case of high amplitude
roughness the whole energetics of the systems changes so dramatically that it
could be impossible to use the experimental results in a meaningful
quantitative way.

The calculation of the neutron count can be done either via the time
dependence of the neutron number at fixed position during the time of flight
$t=L/v_{x}$ or via the decrease of the number of neutrons while they move
along the absorber of the length $L=tv_{x}$. Within the former approach, if
initially there was $N_{j}\left(  0\right)  $ neutrons per unit area in each
discrete state, the time evolution of the densities $N_{j}$ is described by
the particle balance equations $\left(  \ref{n6}\right)  $.

The set of linear differential equations $\left(  \ref{n6}\right)  $ describes
the exponential disappearance of neutron with a set of characteristic
relaxation times $\tau$. These relaxation times $\tau$ are given, according to
Eq. $\left(  \ref{n6}\right)  ,$ by the following characteristic equation:%
\begin{align}
0  &  =\det\left\vert \left(  \frac{1}{\tau}-\frac{1}{\tau_{j}^{\left(
0\right)  }}\right)  \delta_{jj^{\prime}}+\frac{1}{\tau_{jj^{\prime}}%
}\right\vert ,\label{n9}\\
\frac{1}{\tau_{j}^{\left(  0\right)  }}  &  =\frac{1}{v_{j}}%
{\displaystyle\int\nolimits_{\lambda^{\prime}>u_{h}}}
W_{\lambda^{\prime}j}d\lambda^{\prime},\ \frac{1}{\tau_{jj^{\prime}}}=\frac
{1}{v_{j^{\prime}}}W_{jj^{\prime}}-\frac{1}{v_{j}}\delta_{jj^{\prime}}%
\sum_{j^{\prime\prime}\leq Z}W_{jj^{\prime\prime}}\nonumber
\end{align}
where it is convenient to introduce the dimensionless velocities $\beta_{j}$,
$v_{j}=\beta_{j}v_{0},\ v_{0}=\sqrt{2gl_{0}}\sim1.\,\allowbreak0733\times
10^{-2}\ \mathrm{m/s}.$

Eq. $\left(  \ref{n9}\right)  $ yields $\mathcal{N}$ relaxation times
$\tau=\tau_{j}$ and the same number of eigenvectors $\mathbf{N}_{j}$ which
evolve as
\begin{equation}
\mathbf{N}_{j}=\mathbf{N}_{j}\left(  0\right)  \exp\left(  -t/\tau_{j}\right)
. \label{n8}%
\end{equation}
The number of particles in each discrete (gravitational) state $N_{i}$
decreases as%
\begin{equation}
N_{i}=\sum_{j}N_{ij}\left(  0\right)  \exp\left(  -t/\tau_{j}\right)  ,
\label{nnn8}%
\end{equation}
where $N_{ij}$ is the $i$-th component of the $j$-th eigenvector. In other
words, we have to project the initial distribution of neutrons over the
gravitational states, $\left(  N_{1}\left(  0\right)  ,N_{2}\left(  0\right)
,...\right)  $ on the eigenvectors $\mathbf{N}_{j}$. Then the overall neutron
count is%
\begin{equation}
N\left(  h\right)  =\sum_{i,j}N_{ij}\left(  0\right)  \exp\left[  -L_{x}%
/V_{0}\tau_{j}\left(  h\right)  \right]  . \label{n12}%
\end{equation}
where the time is limited by the time of flight between the plates
$t=L_{x}/V_{0}\sim2\times10^{-2}$ \textrm{s}. The dependence $N\left(
h\right)  $ acquires a distinct structure only if there is a pronounced
hierarchy in the set of relaxation times $\tau_{j}\left(  h\right)  $ and if
there are pronounced changes in the \emph{longest }relaxation times $\tau
_{j}\left(  h\right)  $ as a function of the separation between the plates.

It is clear that we should take into account only the relaxation times which
are comparable to or much bigger than the time of flight $t=L_{x}/V_{0}$. For
much shorter relaxation times we do not have to do any precise calculations
and just disregard the components of the initial distribution which are
decaying as the corresponding eigenvectors.

As we will see below, these longest relaxation times $\tau_{j}\left(
h\right)  ,$ which correspond to the particles in the lowest energy states far
away from the absorption edge, experience rather steep changes at certain
values of the spacings between the plates $h_{1}<h_{2}<h_{3}...$ However, this
feature can be observed only for specific types of the roughness of the
scatterer surface when the values $h_{1},h_{2},h_{3}...$ are sufficiently far
away from each other. One of the crucial elements here is the relation between
the times $\tau_{jj^{\prime}}$ and $\tau_{j}^{\left(  0\right)  }$. If
$\tau_{jj^{\prime}}$\ is shorter than $\tau_{j}^{\left(  0\right)  },$ then
the loss of neutrons occur as a result of scattering-driven gradual increase
in gravitational energy (slow diffusion of particles between the states $j$
with a gradual increase in the state index $j$) until the neutron gets into
the absorber. If this is the case, the disappearance times for neutrons from
the neighboring states are comparable and it is not easy to extract the
information on the properties of quantized gravitational states from the
neutron count.

In the opposite case of very short times $\tau_{j}^{\left(  0\right)  }$,
\ the neutron gets absorbed essentially as a result of a single scattering
act. In this case, one can disregard the interstate transitions and $\tau
_{j}\simeq\tau_{j}^{\left(  0\right)  }$. This regime is more advantageous for
experiment especially if there is a distinct hierarchy of times $\tau
_{j}^{\left(  0\right)  }$. The transition between these regimes is determined
by the lateral size (correlation radius) of surface inhomogeneities and the
asymptotic behavior of the correlation function. All this information is
contained in the transition probabilities $W_{jj^{\prime}}\left(
p_{j},p_{j^{\prime}}\right)  $. In what follows we will pay particular
attention to the experimental conditions necessary for the observation of the
proper regime.

Note, that the transitions between the states are largely suppressed,
$\tau_{j\neq j^{\prime}}\gg\tau_{jj}$, when the lateral size of the surface
inhomogeneities on the scatterer $R$ is much larger than the size of quantized
states (in this case, $R\gg l_{0}$, or, in dimensionless units, $r\equiv
R/l_{0}\gg1$) \cite{arm2}. Since the momentum change $\delta\beta$\ as a
result of scattering by inhomogeneities of the size $r$ is of the order of
$\delta\beta\sim1/r$ and is very small for large $r$, it seems that in this
case the interstate transitions are allowed only when the energy gaps between
the states are small. If this were correct, the population of the higher
states would have remained constant no matter how long one would have waited
(the gaps between the states increase with increasing state number $j$
approximately proportionally to $j$). However, this is not completely true
because even at $r\equiv R/l_{0}\gg1$ there are certain values of the spacings
between the plates $h_{j}$ at which the transition channels
$j\longleftrightarrow j+1$ open spontaneously \cite{pon1} leading to a drastic
decrease in the relaxation times: the scattering driven momentum change
$\delta\beta\sim1/r$\ is sufficient for the transition between the states $j$
and $\ j^{\prime}$ when
\begin{equation}
\beta_{j}-\beta_{j+1}\simeq\frac{1}{2}\left(  \lambda_{j+1}-\lambda
_{j}\right)  /\sqrt{e}\simeq1/r \label{n10}%
\end{equation}
or, in normal variables, when the spacing $H$ satisfies the equation
\begin{equation}
\frac{1}{2}\left[  \epsilon_{j+1}\left(  H\right)  -\epsilon_{j}\left(
H\right)  \right]  /\sqrt{Ee_{0}\left(  H\right)  }\simeq l_{0}\left(
H\right)  /R. \label{n11}%
\end{equation}
However, it is highly unlikely that this spontaneous opening of mode coupling
channels affects the neutron count in the existing experimental setups. The
observation of this effect requires drastically different experiments
\cite{yiying1}. To avoid dealing with this issue, we will assume that the size
of inhomogeneities $r$ is relatively small.

In order to find the dynamics of the neutron count, we need information about
the initial distribution of neutrons over the vertical velocities (over
discrete energy states $N_{j}\left(  t=0\right)  $). This distribution
reflects the neutron distribution over the vertical velocities in front of the
slit and is determined by the expansion of the classical wave functions in
front of the slit over the gravitational states. This can, in principle, be
done precisely, but only if one knows the exact distribution of particles over
the vertical velocities in the initial beam. Unfortunately, at present the
distribution of neutrons over the vertical velocities cannot be measured as
accurately as the horizontal distribution. However, even if this distribution
were available, the exact mathematical problem would still have been too
complicated because of the complex structure of the experimental apparatus.

The analysis of experiment, which has been performed in Ref. \cite{prot1},
indicates that the initial distribution is approximately uniform,
\begin{equation}
N_{j}\left(  0\right)  =N\left(  0\right)  /Z\mathrm{,} \label{nnn76}%
\end{equation}
where $N\left(  0\right)  $ is the overall combined density of neutrons in all
gravitational states and $Z$ is the total number of such discrete states
between the plates with $\epsilon_{j}<U_{C}$. Note, that the uniform
distribution $\left(  \ref{nnn76}\right)  $ is a non-equilibrium one and
should evolve in time as a result of scattering by surface inhomogeneities
\textit{even in the absence of the neutron absorption.} The equilibrium
distribution, which does not evolve as a result of non-absorbing scattering
irrespective of the form of collision operator, is given by equation%
\begin{equation}
N_{j}\left(  0\right)  =N\left(  0\right)  \beta_{j}/\left[  Z\sum_{i\leq
Z}\beta_{i}\right]  ,\ N_{j}\left(  0\right)  /\beta_{j}=\mathrm{const}
\label{n7}%
\end{equation}

Luckily for us, the exit neutron count depends mostly only on the initial
population of the lowest states. The lifetime of the neutrons from the higher
states is much shorter and they get absorbed well before getting to the
counter. The velocities $\beta_{i}$ for the lowest states can be considered as
constants, $\beta_{i}\approx\beta_{0}$, and, if there is no singularity in the
velocity distribution in the beam at zero angle, the distribution for the
lowest states is close to uniform, Eq. $\left(  \ref{nnn76}\right)  $, anyway.
In this case, the difference between the distributions $\left(  \ref{n7}%
\right)  $ and $\left(  \ref{nnn76}\right)  $ is negligible as well.

The last remaining issue before solving the transport equations $\left(
\ref{n6}\right)  $ is to relate the transition probabilities $W_{jj^{\prime}}$
to the surface roughness (to the correlation function of surface inhomogeneities).

\section{Scattering probabilities: weak roughness}

Roughness of (one of) the walls leads to scattering of neutrons, which, in
turn, results in the transitions between the states, broadening of the lines,
\textit{etc.} In some intermediate energy range, when the lifetimes are
already short but the spacings between levels are not yet large, this
broadening can even effectively transform the discrete energy levels into a
continuum. However, when the matrix elements are still relatively small, one
can always separate the effects of scattering on the energy spectrum from the
transition probabilities that affect the particle transport \cite{arm2} and,
in our case, the neutron count.

We will start from the case of slight roughness for which there are well
established methods of calculation of the surface-driven scattering
probabilities \cite{arm2} (see also \cite{qq17,r3} and our earlier application
to neutrons in a gravitational trap \cite{local}). Since the geometry of the
problem is somewhat different from our previous applications \cite{arm2,local}%
, the calculation should be modified though the general method remains the same.

The rough scatterer corresponds to the barrier of the height $U_{c}$ with
position $z=H+\xi\left(  x,y\right)  ,$ where $\xi\left(  x,y\right)  $ is the
random function which describes the surface roughness, instead of its "ideal"
position at $z=H$ in Figure 1. If we neglect small gravitational potential
$mgH$ in comparison with the barrier $U_{c}$ (the difference in scales is
$10^{5}$!), the roughness-driven "perturbation" becomes%
\begin{equation}
V\left(  z,x,y\right)  =U_{c}\theta\left(  z-H-\xi\right)  -U_{c}\theta\left(
z-H\right)  \approx-\xi\left(  x,y\right)  \delta\left(  z-H\right)  U_{c}.
\label{nn13}%
\end{equation}

The transition probabilities are given by the squares of the matrix elements
of this perturbation averaged over the surface inhomogeneities,%
\begin{equation}
W_{jj^{\prime}}\left(  \mathbf{q},\mathbf{q}^{\prime}\right)  =\left\langle
\left\vert V_{jpq,j^{\prime}p^{\prime}q^{\prime}}\right\vert ^{2}\right\rangle
_{\xi}. \label{n14}%
\end{equation}
The matrix elements should be calculated using the wave functions
\[
\Psi=\psi\left(  x,y\right)  \psi_{j}\left(  z\right)
\]
where $\psi\left(  x,y\right)  $ are the properly normalized plane waves and
the functions $\psi_{j}\left(  z\right)  $\ are given by Eqs. $\left(
\ref{n1}\right)  -\left(  \ref{nnn55}\right)  $.

The matrix elements of the perturbation $\left(  \ref{nn13}\right)  $ are (cf.
Ref. \cite{arm2})
\begin{align}
V_{jj^{\prime}}  &  \equiv\int\exp\left(  ix\left(  p-p^{\prime}\right)
+iy\left(  q-q^{\prime}\right)  \right)  \xi\left(  x\right)  \psi_{j}\left(
z\right)  \psi_{j^{\prime}}\left(  z\right)  \left[  mgH-U_{c}\right]
\delta\left(  z-H\right)  dxd\mathbf{s}\label{n15}\\
&  =-\xi\left(  p-p^{\prime},q-q^{\prime}\right)  U_{c}\psi_{j}\left(
H\right)  \psi_{j^{\prime}}\left(  H\right)  .\nonumber
\end{align}
The scattering probabilities $\left(  \ref{n14}\right)  $ are then%
\begin{equation}
W_{jj^{\prime}}\left(  p,q;p^{\prime},q^{\prime}\right)  =\left\langle
\left\vert V_{jpq,j^{\prime}p^{\prime}q^{\prime}}\right\vert ^{2}\right\rangle
_{\xi}=\zeta\left(  p-p^{\prime},q-q^{\prime}\right)  U_{c}^{2}\psi_{j}%
^{2}\left(  H\right)  \psi_{j^{\prime}}^{2}\left(  H\right)  \label{n16}%
\end{equation}
where $\zeta\left(  p-p^{\prime},q-q^{\prime}\right)  $ - the so-called power
spectrum of inhomogeneities - is the Fourier image of the correlation function
of surface inhomogeneities $\zeta\left(  x,y\right)  ,$
\begin{equation}
\zeta\left(  x,y\right)  \equiv\left\langle \xi(x_{1},y_{1})\xi(x_{1}%
+x,y_{1}+y)\right\rangle \equiv A^{-1}\int\xi(x_{1},y_{1})\xi(x_{1}%
+x,y_{1}+y)dx_{1}dy_{1}, \label{aa1}%
\end{equation}
and $A$ is the averaging area. Finally, assuming the homogeneity of the
surface in $y$-direction, \emph{i.e.,} the lack of the dependence of
$\zeta\left(  x,y\right)  $ on $y$, we get%
\begin{align}
W_{jj^{\prime}}\left(  p,q;p^{\prime},q^{\prime}\right)   &  =\delta\left(
q-q^{\prime}\right)  W_{jj^{\prime}}\left(  p,p^{\prime}\right)
,\label{n17}\\
W_{jj^{\prime}}\left(  p,p^{\prime}\right)   &  =\zeta\left(  p-p^{\prime
}\right)  U_{c}^{2}\psi_{j}^{2}\left(  H\right)  \psi_{j^{\prime}}^{2}\left(
H\right)  .\nonumber
\end{align}

For computations, we will use the most common Gaussian correlation function,
\begin{equation}
\zeta\left(  x\right)  =\ell^{2}\exp\left(  -x^{2}/2R^{2}\right)
,\ \zeta\left(  p\right)  =\sqrt{2\pi}\ell^{2}R\exp\left(  -p^{2}%
R^{2}/2\right)  , \label{n18}%
\end{equation}
where the amplitude $\ell$\ and the correlation radius $R$ characterize the
averaged height and lateral size of surface inhomogeneities. Note, that though
in many cases the real rough surface can be different from the Gaussian one
\cite{q2,fer1}, there are reasons to believe that the roughness in experiment
\cite{neutron1} is close to Gaussian. \ The results for the relaxation time
can also be sensitive to the type of the asymptotic decay of the power
spectrum of inhomogeneities at large $p$ \cite{pon1}. Below we will choose
such a regime for which the interpretation of experimental data would be the
least sensitive to this uncertainty in parameters of the surface correlator.

In our dimensionless variables, the transition probability $\left(
\ref{n17}\right)  ,\left(  \ref{n18}\right)  $ acquires the form%
\begin{equation}
w_{jj^{\prime}}=\sqrt{2\pi}e_{0}^{2}l_{0}^{3}\eta^{2}r\exp\left[  -\left(
\beta_{j}-\beta_{j^{\prime}}\right)  ^{2}r^{2}/2\right]  u_{c}^{2}\psi_{j}%
^{2}\left(  h\right)  \psi_{j^{\prime}}^{2}\left(  h\right)  , \label{n19}%
\end{equation}
where $\eta=\ell/l_{0}$ and $r=R/l_{0}$. The condition of weak roughness is%
\begin{equation}
\eta\equiv\ell/l_{0}\ll1,r. \label{n20}%
\end{equation}

In experiment, the potential barrier $U_{c}$ is very high and can be
considered infinite in calculations of the roughness-driven scattering
probabilities for transitions between the lower gravitational states. In this
case, Eq. $\left(  \ref{n15}\right)  $\ should be replaced by
\begin{equation}
V_{jj^{\prime}}=-\frac{1}{2m}\xi\left(  p-p^{\prime},q-q^{\prime}\right)
\psi_{j}^{\prime}\left(  H\right)  \psi_{j^{\prime}}^{\prime}\left(  H\right)
\label{n21}%
\end{equation}
and Eq. $\left(  \ref{n19}\right)  $ - by%
\begin{align}
W_{jj^{\prime}}\left(  p_{j}-p_{j^{\prime}}\right)   &  =\frac{\sqrt{2\pi}%
}{4m^{2}}l_{0}^{3}\eta^{2}r\exp\left[  -\left(  \beta_{j}-\beta_{j^{\prime}%
}\right)  ^{2}r^{2}/2\right]  \psi_{j}^{\prime2}\left(  H\right)
\psi_{j^{\prime}}^{\prime2}\left(  H\right) \label{n22}\\
&  =\frac{1}{\tau_{0}}w_{jj^{\prime}},\nonumber
\end{align}
where we introduce the following scale $\tau_{0}$ for transition times:%
\begin{equation}
\frac{1}{\tau_{0}}=\frac{\sqrt{2\pi}}{4m^{2}}\frac{\hbar^{2}}{l_{0}^{3}v_{0}%
}\eta^{2} \label{t1}%
\end{equation}
The form $\left(  \ref{n19}\right)  $ with $U_{c}$ should be used only when
calculating the transitions that involve the upper energy levels and the
continuous spectrum above the threshold.

For the gravitational states $\left(  \ref{nn6}\right)  $, the transitional
probabilities $\left(  \ref{n22}\right)  $ become%
\begin{align}
W_{jj^{\prime}}\left(  p_{j}-p_{j^{\prime}}\right)   &  =\frac{1}{\tau_{0}%
}w_{jj^{\prime}}=\frac{\sqrt{2\pi}}{4m^{2}}\frac{\eta^{2}r}{l_{0}^{3}}%
\exp\left[  -\left(  \beta_{j}-\beta_{j^{\prime}}\right)  ^{2}r^{2}/2\right]
\label{nn22}\\
&  \times a_{j}^{2}a_{j^{\prime}}^{2}\left[  \mathrm{Ai}^{\prime}\left(
h-\overline{\lambda}_{j}\right)  -S\mathrm{Bi}^{\prime}\left(  h-\overline
{\lambda}_{j}\right)  \right]  ^{2}\left[  \mathrm{Ai}^{\prime}\left(
h-\overline{\lambda}_{j^{\prime}}\right)  -S\mathrm{Bi}^{\prime}\left(
h-\overline{\lambda}_{j^{\prime}}\right)  \right]  ^{2}\nonumber
\end{align}
with $A_{j}^{2}$ given by Eq.$\left(  \ref{nn7}\right)  $.

For the square well states $\left(  \ref{nn11}\right)  ,$ the scattering
probabilities acquire the "standard" form \cite{arm2},
\begin{equation}
W_{jj^{\prime}}\left(  p_{j}-p_{j^{\prime}}\right)  =\frac{1}{\tau_{0}%
}w_{jj^{\prime}}=\frac{\sqrt{2\pi}\pi^{4}}{16m^{2}}\frac{\eta^{2}r}{l_{0}^{3}%
}\exp\left[  -\left(  \beta_{j}-\beta_{j^{\prime}}\right)  ^{2}r^{2}/2\right]
\frac{j^{2}j^{\prime2}}{h^{6}}. \label{nn23}%
\end{equation}
We also need the probability of transitions between the gravitational state
$j$ and the square well state $j^{\prime}$:%
\begin{align}
W_{jj^{\prime}}\left(  p_{j}-p_{j^{\prime}}\right)   &  =\frac{1}{\tau_{0}%
}w_{jj^{\prime}}\label{nn24}\\
&  =\frac{\sqrt{2\pi}\pi^{2}}{8m^{2}}\frac{\eta^{2}r}{l_{0}^{3}}\exp\left[
-\left(  \beta_{j}-\beta_{j^{\prime}}\right)  ^{2}r^{2}/2\right]
\frac{j^{\prime2}}{h^{3}}a_{j}^{2}\left[  \mathrm{Ai}^{\prime}\left(
h-\lambda_{j}\right)  -S\mathrm{Bi}^{\prime}\left(  h-\lambda_{j}\right)
\right]  ^{2}.\nonumber
\end{align}

All of the transition probabilities above describe transitions between the
discrete states. In order to complete our calculation of the matrix of the
relaxation times $\left(  \ref{n9}\right)  ,$ we have to get the expressions
for the transitions from the bound states into the continuous states
$1/\tau_{j}^{\left(  0\right)  }$. To avoid extended calculations, we will
first assume that instead of the continuous spectrum we are actually dealing
with particles in a large box $L\gg H$ with a set of very close levels
$k_{n}=\pi n/L$. Then for the transition probability from discrete states $j$
in our well into the "continuous" spectrum inside the box $L$ one should use
Eq. $\left(  \ref{n19}\right)  $,
\begin{align}
\frac{1}{\tau_{j}^{\left(  0\right)  }}  &  =\frac{1}{\tau_{0}\beta_{j}%
}\ u_{c}^{2}l_{0}^{2}r\psi_{j}^{2}\left(  h\right)  \frac{2}{L}\sum
_{\lambda_{n}<e}\frac{\exp\left[  -\left(  \sqrt{e-\lambda_{j}}-\sqrt
{e-\lambda_{n}}\right)  ^{2}r^{2}/2\right]  }{\left[  1+\cos^{2}\left(
h\sqrt{\lambda_{n}}\right)  /\left(  1-u_{c}/\lambda_{n}\right)  \right]
},\label{r3}\\
\lambda_{n}  &  =u_{c}+\left(  \pi nl_{0}/L\right)  ^{2}.\nonumber
\end{align}
Switching back from summation over $k_{n}=\pi n/L$ to the integration over the
continuous energy variable $\lambda$, we get%
\begin{equation}
\frac{1}{\tau_{j}^{\left(  0\right)  }}=\frac{u_{c}^{2}l_{0}r\psi_{j}%
^{2}\left(  h\right)  }{\pi\tau_{0}\beta_{j}}\ \int_{0}^{e-u_{c}}%
\frac{d\lambda}{\sqrt{\lambda-u_{c}}}\frac{\exp\left[  -\left(  \sqrt
{e-\lambda_{j}}-\sqrt{e-u_{c}-\lambda}\right)  ^{2}r^{2}/2\right]  }{1+\left(
1+u_{c}/\lambda\right)  \cos^{2}\left(  h\sqrt{\lambda+u_{c}}\right)  }.
\label{r4}%
\end{equation}

In principle, the above equations for the transition probabilities are
sufficient for solving the transport equations from the previous subsection
and for calculating the absorption rates for neutrons in the channel.

\section{Towards stronger roughness}

It seems much more natural to design an experiment with scatterers with strong
roughness rather than with the slight one: scattering by strong roughness
ensures, almost automatically, a significant turn of the neutron velocity
which results in direct absorption of a neutron by a plate. Therefore,
neutrons form all states, except for the lowest gravitational ones, for which
the wave function does not reach the scatterer, can be absorbed by the plates
as a result of, essentially, a single scattering act. Then the count of the
exiting neutrons will give directly, almost without any ambiguity, the number
of particles in the lowest states that are not scattered. In this case, the
dependence of the neutron count on the width of the gap between the plates
should be the sharpest. If one is interested solely in identification of
quantum states, one should definitely try to ensure that the surface roughness
is strong. However, if one is interested in extracting the quantitative
information from the experiment about the energetics of the system, one should
seriously consider using slight roughness for which the particle dynamics can
be described very accurately as it is shown above.

In other words, if one is interested solely in observing the quantization of
neutrons, one should use a scatterer with strong roughness; if the purpose is
the use of quantized states for ultralow energy calibration, one should use
only rough systems which are subjects to a quantitative theory. Therefore, it
is important to discuss the restrictions imposed by the conditions of slight
roughness and the possibility of extending the limits of applicability of our
theory \textit{in this particular case }to stronger roughness. There are two
types of strong roughness: high-amplitude and high-aperture roughness. As we
will see, our theory can be extended to the high-aperture low-amplitude
roughness, but not to the high-amplitude roughness. In the last Section we
will argue that the high-amplitude roughness should not be used in experiment either.

The surface roughness is considered to be strong if one of the conditions of
slight roughness,%

\begin{equation}
\eta\equiv\ell/l_{0}\ll1,r \label{n23}%
\end{equation}
breaks down. In such cases, one cannot usually say much except that one
surface collision is sufficient for a complete dephasing of a particle. In our
problem such an outcome could even have been looked at as beneficial since
this would immediately allow us to estimate the corresponding relaxation time
approximately as
\begin{equation}
\tau_{j}=H/v_{zj}. \label{n24}%
\end{equation}
However, such an assertion is not always correct since the scattering
probability $W$ depends not only on the parameters of the surface (the
correlation function of surface roughness $\zeta$) but also on the probability
of the particles to get close to the wall which enters $W$ via the particle
wave functions and/or their derivatives on the wall (see, \emph{e.g., }Eqs.
$\left(  \ref{n19}\right)  ,\left(  \ref{n22}\right)  $ for $W$ for weak
roughness). When such a probability is low, one can (and should) sometimes
deal with the strong roughness as if it is weak.

The calculation of the scattering probability consists of two steps:
calculation of the matrix elements of the roughness-driven "perturbation"
$V_{jj^{\prime}}$ and the calculation of the scattering probabilities $W$
using these matrix elements. Since the latter step is easier to analyze, we
will start from it.

This second step is also less restrictive. Here one has to express the
scattering probability $W$ via the matrix elements $V_{jj^{\prime}}$. The
scattering probability is always given by the expression of the type $\left(
\ref{n14}\right)  ,$\ but with the scattering $T$-matrix $\widehat{T}$ instead
of the potential distortion $\widehat{V}$. The link between these two matrices
is given by the operator equation%
\begin{equation}
\widehat{T}=\widehat{V}+\widehat{T}\widehat{G}\widehat{V}\label{n28}%
\end{equation}
where $\widehat{G}$ is the Green's function. For quantized gravitational-like
states $j,j^{\prime}$\ (the lower states with $h-\lambda_{j,j^{\prime}}\gg1$),
the position of the wall is deep under the barrier where the wave functions
are attenuating exponentially, roughly as $\exp\left[  -\left(  2/3\right)
\left(  h-\lambda\right)  ^{3/2}\right]  $. Then the matrix elements of
$\widehat{V}$ are exponentially small and Eq. $\left(  \ref{n28}\right)  $
leads to $\widehat{T}\approx\widehat{V}$ resulting in the validation of Eq.
$\left(  \ref{n14}\right)  $ even when the distortion $V$, by itself, is not
small. This means that Eq. $\left(  \ref{n14}\right)  $ can be used as long as
we are interested solely in the transitions to and from the gravitational
states for which the matrix elements are always small even for not very small
roughness. One should be much more careful with the transitions between the
higher states.

The simplest way of evaluating the matrix elements $V_{jj^{\prime}}$ seems to
be a generalization of approach \cite{qq17} (see also the first Ref.
\cite{arm2}) for the roughness-driven perturbation%
\begin{equation}
V\left(  z,x,y\right)  =U_{c}\theta\left(  z-H-\xi\right)  -U_{c}\theta\left(
z-H\right)  . \label{n25}%
\end{equation}
Instead of using Eq. $\left(  \ref{nn13}\right)  $, we now write $V$ as an
expansion%
\begin{equation}
V\left(  z,x,y\right)  =U_{c}\sum_{1}^{\infty}\frac{\left(  -1\right)  ^{n}%
}{n!}\delta^{\left(  n-1\right)  }\left(  z-H\right)  \xi^{n}\left(
x,y\right)  \label{nn25}%
\end{equation}
with the matrix elements%
\begin{align}
V_{jj^{\prime}}\left(  \mathbf{q-q}^{\prime}\right)   &  =U_{c}\sum
_{1}^{\infty}\frac{\left(  -1\right)  ^{n}}{n!}\int_{0}^{H}\psi_{j}\left(
z\right)  \delta^{\left(  n-1\right)  }\left(  z-H\right)  \psi_{j^{\prime}%
}\left(  z\right)  dz\label{nnn25}\\
&  \times\int\xi^{n}\left(  \mathbf{s}\right)  \exp\left[  i\left(
\mathbf{q-q}^{\prime}\right)  \cdot\mathbf{s}\right]  d\mathbf{s.}\nonumber
\end{align}
The first integral reduces to a set of higher derivatives of the wave
functions on the wall, $\psi_{j}^{\left(  m\right)  }\left(  H\right)
\psi_{j^{\prime}}^{\left(  n-1-m\right)  }\left(  H\right)  $, or, in
dimensionless variables, to $l_{0}^{-n+1}\psi_{j}^{\left(  m\right)  }\left(
h\right)  \psi_{j^{\prime}}^{\left(  n-1-m\right)  }\left(  h\right)  $.
Therefore, if the amplitude of the corrugation is $\ell$, the expansion
$\left(  \ref{nnn25}\right)  $\ becomes an expansion in $\ell/l_{0}$. If the
wave functions on the walls are (exponentially) small, as it is for the lowest
gravitational states, the matrix elements $V_{jj^{\prime}}$ can remain small
even for not very small values of the corrugation amplitude $\ell$. However,
in this case the all the constants in the expressions for the scattering
probabilities change: instead of the wave functions on the walls one should
write a proper combinations of $\psi_{j}^{\left(  m\right)  }\left(  h\right)
$ from Eq. $\left(  \ref{nnn25}\right)  $ and, more importantly, the
correlation function of the surface corrugation should be replaced by the
Fourier images of the higher moments of the correlation.

If, on the other hand, the amplitude of the corrugation is small, $\ell\ll
l_{0}$, Eq. $\left(  \ref{nnn25}\right)  $ leads to exactly the same
expression for the matrix elements as Eqs. $\left(  \ref{n15}\right)  $\ and,
therefore, to the scattering probabilities $\left(  \ref{n14}\right)
$,$\left(  \ref{n19}\right)  $,$\left(  \ref{n22}\right)  $ \textit{even when
the aperture of the inhomogeneities is large.} This means that, as long as the
matrices $\widehat{V}$ are small and are close to the $\widehat{T}$-matrix,
$\widehat{V}\approx\widehat{T}$ , the high-aperture roughness could still be
treated in the same way as weak roughness while the high-amplitude roughness cannot.

The above analysis on the basis of a generalized approach \cite{qq17} is not
completely accurate. A a more accurate analysis can be done using the mapping
transformation method (see both references \cite{arm2}), which involves the
lateral derivatives of the surface roughness in a more natural way than the
method of Ref. \cite{qq17}. This method involves the first lateral derivatives
$\xi_{x,y}^{\left(  1\right)  }$, which have the order of magnitude $\ell/R$
and not $\ell/l_{0}$, explicitly. As a result, it looks that the expressions
for the matrix elements $V_{jj^{\prime}}$ for the high-aperture roughness are
different from Eq. $\left(  \ref{n15}\right)  .$\ However, the collision
operator in the transport equation involves not all the matrix elements of
$V_{jj^{\prime}}\left(  \mathbf{q,q}^{\prime}\right)  ,$ but, since the
scattering probabilities are always accompanied by the energy $\delta
$-functions $\delta\left(  E-E^{\prime}\right)  $, only those that satisfy the
energy conservation law. This means that Eq. $\left(  \ref{n14}\right)  $ can
be rewritten for our purposes as
\begin{equation}
W_{jj^{\prime}}\left(  \mathbf{q},\mathbf{q}^{\prime}\right)  \delta\left(
\epsilon_{j}+\frac{q^{2}}{2m}-\epsilon_{j^{\prime}}-\frac{q^{\prime2}}%
{2m}\right)  =\left\langle \left\vert V_{jpq,j^{\prime}p^{\prime}q^{\prime}%
}\right\vert ^{2}\right\rangle _{\xi}\delta\left(  \epsilon_{j}+\frac{q^{2}%
}{2m}-\epsilon_{j^{\prime}}-\frac{q^{\prime2}}{2m}\right)  . \label{nn27}%
\end{equation}
It is relatively easy to see \cite{arm2} that the terms with the derivatives
of $\xi$ contribute only to the matrix elements with different energies $E$
and $E^{\prime}$ if the amplitude of roughness is small, $\ell\ll l_{0}$.
Therefore, the expressions for the matrix elements at $E=E^{\prime}$ keep
their weak roughness form even for the strong roughness with high aperture,
but with low amplitude.

Summarizing, the scattering probabilities remain the same as long as the
\textit{amplitude} of roughness $\eta=\ell/l_{0}$ is small, while the
\textit{aperture }of roughness $\eta/r=\ell/R$ can become large,
\begin{equation}
R\ll\ell\ll l_{0}. \label{n27}%
\end{equation}
This means that Eqs. $\left(  \ref{n19}\right)  $,$\left(  \ref{n22}\right)
$\ for the scattering probabilities are valid not only for the weak roughness
$\left(  \ref{n20}\right)  ,\left(  \ref{n23}\right)  ,$\ but also for a much
stronger high-aperture roughness $\left(  \ref{n27}\right)  ,$ \emph{but only
to the extent that one is interested in the transitions to and from the
gravitational states with exponentially small wave functions on the rough
wall}.\ In this case one can often disregard the correlation exponents
$\exp\left[  -\left(  \beta_{j}-\beta_{j^{\prime}}\right)  ^{2}r^{2}/2\right]
$ in the expressions for $W$ and make the rest of the calculations as if the
roughness were weak. Physically the disappearance of these exponents means
that the changes of momenta in scattering are now unrestricted. However, the
moment the transitions between the higher states come into play, these
equations loose their accuracy.

On the other hand, the large \textit{amplitude} of roughness always leads to a
completely different picture than the weak roughness. Note, that if we are
interested in identification of the gravitational states when the distance
between the walls $H$ is comparable to the level size $l_{0}$, $H\gtrsim$
$l_{0}$, the condition $\ell\ll l_{0}$ is a necessary condition for the
\textit{existence} of the gravitational states. Therefore, we have to assume
that this condition holds and that the only important type of strong roughness
is the one given by Eq. $\left(  \ref{n27}\right)  $. In the case of
high-amplitude roughness with $\ell\gg l_{0},$ Eq. $\left(  \ref{n14}\right)
$\ still holds \textit{for the gravitational states} and the structure of Eqs.
$\left(  \ref{n19}\right)  $,$\left(  \ref{n22}\right)  $\ remains the same,
though, instead of the correlation function of surface inhomogeneities
$\zeta\left(  \pi_{j}-\pi_{j^{\prime}}\right)  $,\ the expressions for $W$
contain the summation over the higher order correlators. However, as it was
explained above, we are not interested in this situation. What is more, under
these conditions the state energies experience such large roughness-driven
changes that the classification of states on the basis of an ideal well looses
all meaning.

For practical purposes, the results of this section mean that the our theory
is still valid for a relatively strong roughness $\left(  \ref{n27}\right)  $,
but only in application to the lowest gravitational states and not to the
square wall states. Since in this paper we were interested mostly in the
direct transitions from the gravitational states into continuum, these results
still hold for $1\gg\eta\gg r$ and the results in Figure 11 below are still
within the applicability of the theory and the step-wise depletion of neutrons
can be observed.

\section{Relaxation times}

\subsection{Direct absorption \textit{vs. }protracted diffusion between the
states}

There are two ways how the neutrons can be absorbed by the plates (transit
into the continuous spectrum) as a result scattering-driven interstate
transitions. The first way is a relatively protracted diffusion between the
states which results, eventually, in neutron getting into continuum. The
second way is by direct transitions from the initial state into the continuum
without involving any intermediate states. For the latter method to dominate,
the probabilities of direct transitions into continuum should dominate over
transitions between individual discrete states. Obviously, this direct process
has an undeniable theoretical appeal since in this case we do not need to
solve the set of $120h\times120h$\ transport equations $\left(  \ref{n6}%
\right)  $ and only calculate the transition times for direct transitions into
continuum $\tau_{j}^{\left(  0\right)  }$.

However, as we will see below, following this route has important advantages
for experiment as well. If the direct transitions dominate, the results will
be the least sensitive to the correlation characteristics of the surface
roughness (both to the correlation parameters and to the shape of the
correlation function) \emph{and} to the details of the initial distribution of
neutrons over the energy levels. Then the interpretation of experimental data
should be more reliable and unambiguous than is the case of protracted
interstate diffusion.

We will start from delineating conditions under which the direct transitions
into continuum from all discrete states, including the lowest, are possible
and dominant. The next step is the calculation of the corresponding transition
rates which will immediately lead us to the results for the neutron count.

The most relevant time scale is the time of flight
\begin{equation}
t=L_{x}/V_{0}\sim2\times10^{-2}\mathrm{s}. \label{r0}%
\end{equation}
Numerically, the scale of the relaxation times in $\left(  \ref{n9}\right)  $
is given, according to Eqs. $\left(  \ref{n22}\right)  $-$\left(
\ref{nn24}\right)  ,$ by the coefficient $\tau_{0}$ $\left(  \ref{t1}\right)
$ in the scattering probabilities $W,$
\begin{equation}
\tau_{0}^{-1}=\frac{\sqrt{2\pi}}{4m^{2}}\frac{\hbar^{2}}{l_{0}^{3}v_{0}}%
\eta^{2}\sim1.\,\allowbreak15\times10^{3}\eta^{2}\ \mathrm{s}^{-1} \label{r1}%
\end{equation}
where $\eta=\ell/l_{0}$ is the dimensionless amplitude of the surface
corrugation, and $t/\tau_{0}\sim23\eta^{2}/\beta$. Since $l_{0}\sim$ $5.871$
$\mu\mathrm{m}$ and the spacing between the walls in experiment does not
exceed $50$ $\mu\mathrm{m}$, the reasonable limits on the amplitude of
inhomogeneities are $10^{-3}\lesssim\eta=\ell/l_{0}\lesssim0.1,$ and%
\begin{equation}
2.3\times10^{-5}\lesssim t/\tau_{0}\lesssim0.23. \label{r2}%
\end{equation}
(in existing experiments, $0.03\lesssim\eta\lesssim0.1$ and the lower limit in
the inequality $\left(  \ref{r2}\right)  $ is about $0.02$).

The full relaxation times differ from $\tau_{0}^{-1}$ by the dimensionless
factors $w$. To simplify the problem, we have to follow the largest of these
$w$. Note, that according to the above expressions for $W$ (see,
\textit{e.g.,} Eq. $\left(  \ref{nn23}\right)  $) the most probable
transitions are the transitions into the \textit{highest allowed} (by the
correlation exponent) states. Therefore, the direct transitions into the
continuous spectrum, \textit{when allowed}, have the shortest transition
times. What is more, the corresponding inverse times $\tau_{j}^{\left(
0\right)  -1}$ contain summation over all the accessible states in the
continuum and could become noticeably shorter than the transition times
between the individual discrete states.

There are two types of factors entering the dimensionless probabilities $w$:
the correlation exponents with the negative indices $-\left(  \beta_{j}%
-\beta_{j^{\prime}}\right)  ^{2}r^{2}/2,$ which can only decrease $w$, and the
factors related to the values of the wave functions on the walls. The
correlation exponents, in general, encourage the transitions with the smallest
change in momentum, \emph{i.e., }the transitions between the nearby states,
unless, of course, the correlation length $r$ is small (see below) and it does
not matter. The coefficients originating from the wave functions on the wall,
on the other hand, favor the direct transitions into (and from) the highest
states (factors $j^{\prime2}/h^{3}$\ in Eqs. $\left(  \ref{nn23}\right)
,\left(  \ref{nn24}\right)  $). The balance of these opposing factors
determines whether the attenuation of the gravitational states occurs via the
gradual diffusion between the states or via direct transitions into the
highest states.

The possibility of the transitions is determined by the momentum transfer in
the exponent of the correlation function,
\begin{equation}
\left(  \beta_{j}-\beta_{j^{\prime}}\right)  ^{2}r^{2}/2=\left(
\sqrt{e-\lambda_{j}}-\sqrt{e-\lambda_{j^{\prime}}}\right)  ^{2}r^{2}%
/2.\label{r11}%
\end{equation}
This means that the direct transitions from the lowest states into the
continuous spectrum (and all transition between the discrete states) are
\emph{not }suppressed if%
\begin{equation}
\left(  \sqrt{1/\chi}-\sqrt{1/\chi-1}\right)  ^{2}u_{c}r^{2}/2\lesssim
1,\label{k2}%
\end{equation}
($\chi\equiv u_{c}/e$) or, using the numeric value for $u_{c}$,%
\begin{equation}
r\lesssim\allowbreak3.\,\allowbreak8\times10^{-3}F_{0}\left(  \chi\right)
,\ R\lesssim2.\,\allowbreak227\,0\times10^{-2}F_{0}\left(  \chi\right)
\ \mathrm{\mu m}\text{, }F_{0}\left(  x\right)  =\sqrt{x}/\left(  1-\sqrt
{1-x}\right)  \label{k3}%
\end{equation}
where the function $F_{0}\left(  \chi\right)  $ is given in Figure 2.%

\begin{figure}
[ptb]
\begin{center}
\includegraphics[
natheight=4.323200in,
natwidth=5.833200in,
height=3.8095in,
width=5.1318in
]%
{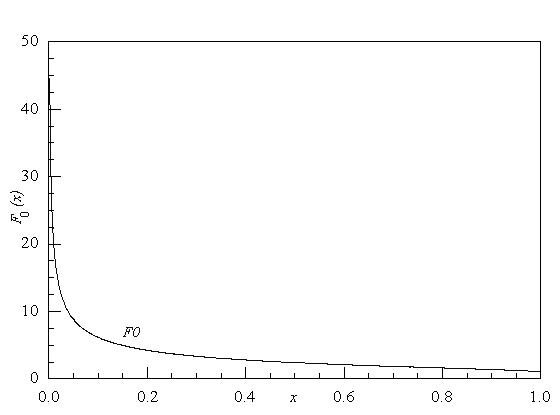}%
\caption{Function $F_{0}\left(  x\right)  $, Eq. $\left(  \ref{k3}\right)  $.}%
\end{center}
\end{figure}

This condition is not overly restrictive. Unless the overall velocity of the
beam is close to the absorption threshold, \textit{i.e., }unless $\chi
=u_{c}/e$ is close to $1$, the condition $\left(  \ref{k3}\right)  $ can
easily be satisfied. Essentially, condition $\left(  \ref{k3}\right)
$\ determines whether the particles, which are initially in the lower discrete
states, can disappear as a result of direct transitions into continuous
spectrum (or into short-lived upper states).

Since the transitions between the lower states are much less likely than the
transitions between the lower and higher states and between the higher states,
we can use expressions $\left(  \ref{nn24}\right)  $ for all $\tau
_{jj^{\prime}}^{-1}$ for the gravitational states and Eq. $\left(
\ref{nn23}\right)  $ for all square well states.

\subsection{Transitions from the square well states into continuous spectrum
(direct absorption)}

We should start from estimating the times $\tau_{j}^{\left(  0\right)  }$ for
transitions into continuous spectrum $\left(  \ref{r4}\right)  $: if these
times are short, at least for some discrete states $j$, $t/\tau_{j}^{\left(
0\right)  }\gg1$, this will help in evaluating the relaxation times for all
other states as well. First, since $u_{c}$ is large, one can replace $\cos
^{2}\left(  h\sqrt{\lambda+u_{c}}\right)  $ in the denominator of Eq. $\left(
\ref{r4}\right)  $ by $1/2$,
\begin{equation}
\frac{1}{\tau_{j}^{\left(  0\right)  }}=\frac{u_{c}^{2}l_{0}r\psi_{j}%
^{2}\left(  h\right)  }{\pi\tau_{0}\beta_{j}}\ \int_{0}^{e-u_{c}}%
\frac{d\lambda}{\sqrt{\lambda+u_{c}}}\frac{\exp\left[  -\left(  \sqrt
{e-\lambda_{j}}-\sqrt{e-u_{c}-\lambda}\right)  ^{2}r^{2}/2\right]  }{1+\left(
1+u_{c}/\lambda\right)  /2}.\label{r5}%
\end{equation}
Since $\lambda_{j}<u_{c}<e$, the index in the exponent changes from $\left(
\sqrt{e-\lambda_{j}}-\sqrt{e-u_{c}}\right)  ^{2}r^{2}/2$ \ \ to $\left(
e-\lambda_{j}\right)  r^{2}/2>\left(  e-u_{c}\right)  r^{2}/2\sim10^{5}r^{2}$.
If $r$ is small, $r\lesssim10^{-3}$ ($R\lesssim5$ \textrm{nm}), as suggested
in the previous section, the index in the exponent is always small and the
correlation exponent can be disregarded,
\begin{align}
\frac{1}{\tau_{j}^{\left(  0\right)  }} &  =\frac{u_{c}^{2}l_{0}r\psi_{j}%
^{2}\left(  h\right)  }{\pi\tau_{0}\beta_{j}}\ \int_{0}^{e-u_{c}}%
\frac{d\lambda}{\sqrt{\lambda+u_{c}}}\frac{2}{3+u_{c}/\lambda}=\frac
{u_{c}^{5/2}l_{0}r\psi_{j}^{2}\left(  h\right)  }{\pi\tau_{0}\beta_{j}}%
2\int_{0}^{1/\chi-1}dz\frac{z}{3z+1}\frac{1}{\sqrt{1+z}}\label{r8}\\
&  \approx10\frac{u_{c}^{5/2}l_{0}r\psi_{j}^{2}\left(  h\right)  }{\pi\tau
_{0}\beta_{j}}F_{1}\left(  \chi\right)  ,\ F_{1}\left(  x\right)
\approx0.11\left[  1.2/\sqrt{x}-1.76+0.245\ln\left(  \frac{6}{3-2.45\sqrt{x}%
}-1\right)  \right]  .\nonumber
\end{align}
The function $F_{1}\left(  x\right)  $, is plotted in Figure 3.%

\begin{figure}
[ptb]
\begin{center}
\includegraphics[
natheight=4.114800in,
natwidth=6.146200in,
height=3.8009in,
width=5.6645in
]%
{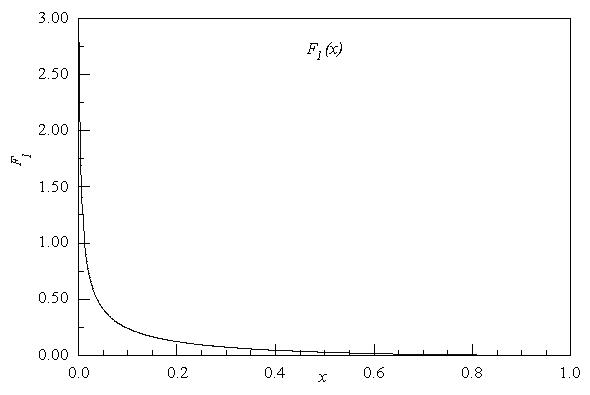}%
\caption{Function $F_{1}\left(  x\right)  $, Eq. $\left(  \ref{r8}\right)  $.}%
\end{center}
\end{figure}

For the estimates, the function $F_{1}\left(  \chi\right)  $ is mostly
irrelevant except for $e\rightarrow u_{c}$ $\left(  \chi\rightarrow1\right)  $
. According to Eq. $\left(  \ref{nnn19}\right)  $, for deep square-well levels%
\begin{equation}
\frac{1}{\tau_{j}^{\left(  0\right)  }}\sim\frac{4.6\times10^{13}r}{\beta
_{j}\tau_{0}}\frac{2\overline{\lambda_{j}}}{hu_{c}}F_{1}\left(  \chi\right)
.\label{r9}%
\end{equation}
(for higher levels, one should use for $\psi_{j}^{2}\left(  h\right)  $ a more
accurate Eq. $\left(  \ref{nnn17}\right)  $).

Taking into account Eqs. $\left(  \ref{r0}\right)  ,\left(  \ref{r1}\right)  $
for the time of flight $t$ and characteristic time $\tau_{0}$, it is
convenient to represent the ratio of the time of flight to the absorption time
as
\begin{equation}
\frac{t}{\tau_{j}^{\left(  0\right)  }}\sim10^{4}\eta^{2}\frac{(10^{3}%
r)}{h\left(  10^{-3}\beta_{j}\right)  }\left(  10^{5}\lambda_{j}/u_{c}\right)
F_{1}\left(  \chi\right)  . \label{r10}%
\end{equation}
Since even for the lowest square-well states $\lambda_{j}/u_{c}>10^{-5},$ this
means that neutrons from all the square-well states, even for the lowest ones,
disappear directly into the continuous spectrum during the time of flight at
any reasonable amplitude of roughness and small lateral size of
inhomogeneities as long as the correlation radius of surface roughness $r$ is
sufficiently small and the absorption threshold is not close to the overall
kinetic energy (the ratio $\chi=u_{c}/e$ is not close to 1).

At larger size of inhomogeneities, the restrictions on the momentum transfer
become more important and the integral in Eq. $\left(  \ref{r5}\right)  $
should be evaluated more carefully. For the lower states $\lambda_{j}\ll e,$
the absorption time is given by Eq. $\left(  \ref{r8}\right)  $ in which the
integral $F_{1}\left(  \chi\right)  $ should be replaced by the function
$F\left(  \chi,\sqrt{u_{c}}r\right)  :$
\begin{align}
\frac{1}{\tau_{j}^{\left(  0\right)  }} &  =10\frac{u_{c}^{5/2}l_{0}r\psi
_{j}^{2}\left(  h\right)  }{\pi\tau_{0}\beta_{j}}\ F\left(  \chi,\sqrt{u_{c}%
}r\right)  ,\ \label{r12}\\
F\left(  x,y\right)   &  =0.2\int_{0}^{1/x-1}dz\frac{z}{3z+1}\frac{1}%
{\sqrt{1+z}}\exp\left[  -\left(  \sqrt{1/x}-\sqrt{1/x-1-z}\right)  ^{2}%
y^{2}/2\right]  .\nonumber
\end{align}
The plots of the function $F\left(  \chi,\sqrt{u_{c}}r\right)  $ at two values
of $r,$ $r=0.01;0.001$ are given in the Figure 4. The same function $F\left(
\chi,\sqrt{u_{c}}r\right)  $ is plotted as a function of $r$ in Figure 5 at
$\chi=0.15;0.3;0.5$. As it can be seen in the plots, choosing proper values
for the correlation radius of inhomogeneities $r$ and the ratio of the
threshold energy to the overall kinetic energy is crucially important for
having effective absorption.%

\begin{figure}
[ptb]
\begin{center}
\includegraphics[
natheight=4.375100in,
natwidth=6.167000in,
height=4.011in,
width=5.6438in
]%
{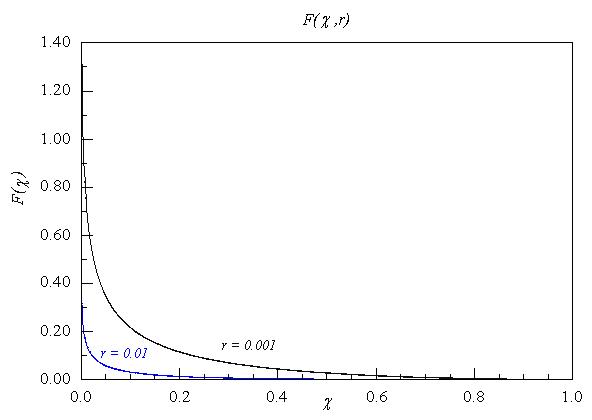}%
\caption{Function $F\left(  \chi,\sqrt{u_{c}}r\right)  $, Eq. $\left(
\ref{r12}\right)  $, at three values of $\chi,$ $\chi=0.15;0.3;0.5$.}%
\end{center}
\end{figure}
%

\begin{figure}
[ptb]
\begin{center}
\includegraphics[
natheight=4.416600in,
natwidth=6.114200in,
height=4.1597in,
width=5.7484in
]%
{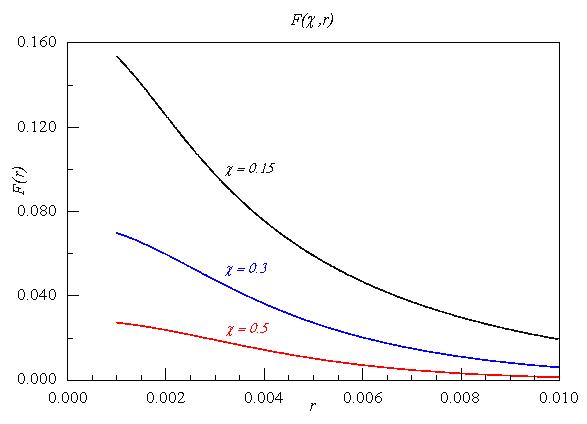}%
\caption{Function $F\left(  \chi,\sqrt{u_{c}}r\right)  $, Eq. $\left(
\ref{r12}\right)  $, at three values of $\chi,$ $\chi=0.15;0.3;0.5$.}%
\end{center}
\end{figure}

Function $F\left(  \chi,\sqrt{u_{c}}r\right)  $ decreases rapidly with
increasing $\chi$ and even more rapidly with increasing $r$. Since
$\sqrt{u_{c}}\sim370$ and the integrand depends exponentially on $\sqrt{u_{c}%
}r$, the dependence of $F\left(  \chi,\sqrt{u_{c}}r\right)  $ on $r$ is very
steep: when $r>0.03$ the momentum transfer restrictions are so severe that the
direct transitions from the lower levels to the continuous spectrum are
suppressed by orders of magnitude unless $\chi$ is very small [in existing
experiment, $\chi\gtrsim0.15$]. For example, when $r=0.01$ the function
$F\left(  \chi,\sqrt{u_{c}}r\right)  $ reaches $10^{-2}$ only at $\chi
\simeq0.26$.

With the same numbers as in Eq. $\left(  \ref{r10}\right)  $, the ratio of the
time of flight to the time of direct transitions into the continuous spectrum
is
\begin{equation}
\frac{t}{\tau_{j}^{\left(  0\right)  }}\sim10^{4}\eta^{2}\frac{(10^{3}%
r)}{h\left(  10^{-3}\beta_{j}\right)  }\left(  10^{5}\lambda_{j}/u_{c}\right)
F\left(  \chi,\sqrt{u_{c}}r\right)  . \label{r13}%
\end{equation}

\subsection{Direct transitions from the gravitational states into the
continuous spectrum}

Probabilities of direct transitions from the gravitational states into
continuous spectrum are given by the same Eq. $\left(  \ref{r5}\right)  $ $,$
but with different values of $\psi_{j}\left(  h\right)  $, Eq. $\left(
\ref{nn14}\right)  $ instead of Eq. $\left(  \ref{nnn17}\right)  $. Since the
integral remains the same, the only substantial change is the replacement of
$\psi\left(  h\right)  $\ \ in the \emph{r.h.s.} of Eq. $\left(
\ref{r12}\right)  $ by $B_{j}$ $\left(  \ref{nn14}\right)  ,$
\begin{align}
\frac{t}{\tau_{j}^{\left(  0\right)  }}  &  \sim10^{4}\eta^{2}\frac{(10^{3}%
r)}{10^{-3}\beta_{j}}b_{j}F\left(  \chi,\sqrt{u_{c}}r\right)  ,\label{r14}\\
b_{j}  &  \equiv10^{5}l_{0}B_{j}^{2}/2=10^{5}a_{j}\left[  \mathrm{Ai}^{\prime
}\left(  h-\overline{\lambda}_{j}\right)  -\overline{S}_{j}\mathrm{Bi}%
^{\prime}\left(  h-\overline{\lambda}_{j}\right)  \right]  ^{2}/u_{c}%
\nonumber\\
&  \simeq0.3a_{j}\left[  \mathrm{Ai}^{\prime}\left(  h-\overline{\lambda}%
_{j}\right)  -\overline{S}_{j}\mathrm{Bi}^{\prime}\left(  h-\overline{\lambda
}_{j}\right)  \right]  ^{2}\nonumber
\end{align}
where $a_{j}$ is given by Eq. $\left(  \ref{nn7}\right)  $,%
\begin{equation}
a_{j}=\left\{  \left(  \mathrm{Ai}^{\prime}\left(  -\overline{\lambda}%
_{j}\right)  -S\mathrm{Bi}^{\prime}\left(  -\overline{\lambda}_{j}\right)
\right)  ^{2}-\left(  \mathrm{Ai}^{\prime}\left(  h-\overline{\lambda}%
_{j}\right)  -\overline{S}\mathrm{Bi}^{\prime}\left(  h-\overline{\lambda}%
_{j}\right)  \right)  ^{2}\right\}  ^{-1}.\nonumber
\end{equation}

This means that the neutron in the gravitational states disappear on the time
of flight unless the coefficient $a_{j}\left[  \mathrm{Ai}^{\prime}\left(
h-\overline{\lambda}_{j}\right)  -\overline{S}_{j}\mathrm{Bi}^{\prime}\left(
h-\overline{\lambda}_{j}\right)  \right]  ^{2}$ and/or the function $F\left(
\chi,\sqrt{u_{c}}r\right)  $ are very small. This is so when the distance
between the walls $h$ significantly exceeds $\overline{\lambda}_{j}$ and the
wave function on the wall and, therefore, $b_{j}$ are exponentially small. To
evaluate this effect we need to calculate numerically the eigenvalues
$\overline{\lambda}_{j}\left(  h\right)  $, Eq. $\left(  \ref{nn6}\right)  $
and the coefficients $b_{j}$, Eq. $\left(  \ref{r14}\right)  $.

The dependence of the nine lowest (gravitational) energy levels on the spacing
is presented in Figure 6. The crosses on the curves mark the points where the
"size" of the level becomes equal to the distance between the walls (in our
notations, $\lambda=h$) and the wave functions start "touching" the walls or,
in other words, the levels move from the gravitational to the square well domain.%

\begin{figure}
[ptb]
\begin{center}
\includegraphics[
natheight=4.228900in,
natwidth=6.010500in,
height=3.736in,
width=5.3004in
]%
{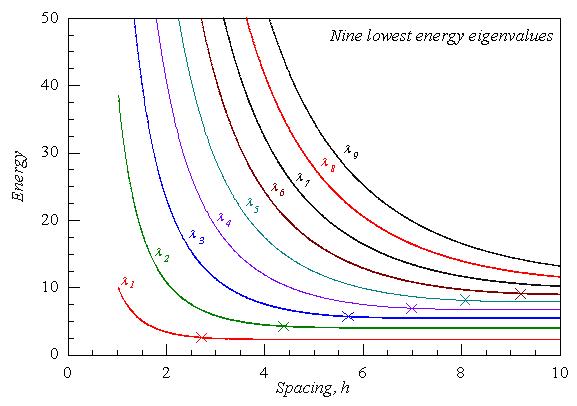}%
\caption{Nine lowest energy levels $\lambda_{i}$ as a function of the spacing
between the walls $h$. The stars mark the positions where the size of the
energy of the level becomes equal to $mgH$ (in dimensionless
variables,\ $\lambda_{i}=h$) and the level shifts to the "square well"
domain.}%
\end{center}
\end{figure}

The coefficients $b_{j}\left(  h\right)  ,$ Eq. $\left(  \ref{r14}\right)  $,
are presented in Figures 7,8 in two different scales. As we can see, the
curves are extremely steep. This means that when scanning the distance between
the walls $h$ one encounters the critical values $h_{j}$ at which $t/\tau
_{j}^{\left(  0\right)  }$ changes from very small to very large values.
According to numerical data, $b_{j}\left(  h\right)  $ changes between
$10^{-4}$ and $10^{-2}$ when $5.87>h>4.13$ for $b_{1}$, $6.64>h>5.76$ for
$b_{2}$, $8.62>h>7.12$ for $b_{3}$, $10>h>8.34$ for $b_{4}$, \textit{etc.} The
change of $b$ for all these levels from $10^{-4}$ to $10^{-3}$\ requires
change in $h$ by only 0.2. Since such change in $h$ is equivalent to the
change in the distance between the walls by about $1.1$ $\mathrm{\mu m}$
\emph{and} the neutron count depends exponentially on the coefficients $b,$
the neutron count should, under proper circumstances, become in experiment a
truly step-wise function of the clearance between the walls. However, since
the functions $b\left(  h\right)  $ for higher levels overhang over each
other, it is rather unlikely if within this observation technique one can
resolve more than three or four gravitational levels.%

\begin{figure}
[ptb]
\begin{center}
\includegraphics[
natheight=4.427000in,
natwidth=5.760500in,
height=4.1511in,
width=5.3921in
]%
{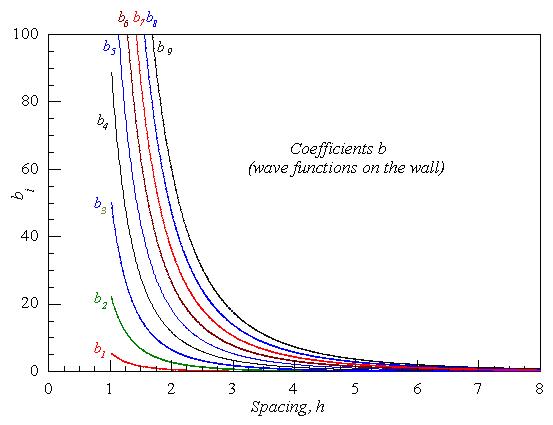}%
\caption{The coefficients $b_{j}\left(  h\right)  ,$ Eq. $\left(
\ref{r14}\right)  $, which determine the square of the wave function on the
surface of the scatterer, $B_{j}^{2}$.}%
\end{center}
\end{figure}
%

\begin{figure}
[ptb]
\begin{center}
\includegraphics[
natheight=4.812700in,
natwidth=6.072700in,
height=3.9816in,
width=5.0194in
]%
{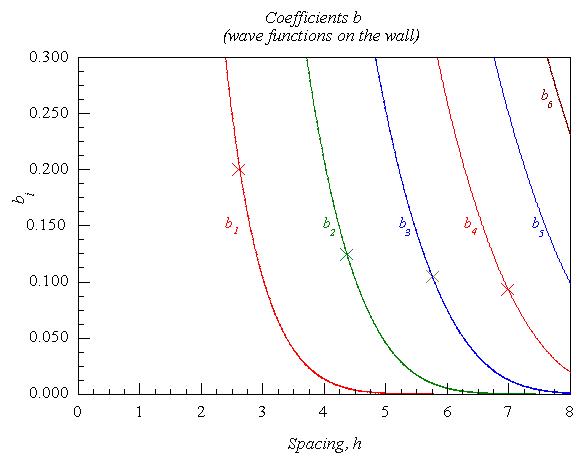}%
\caption{The same as in Figure 7 but on a smaller scale. The stars on the
curves mark the same points as in Figure 6.}%
\end{center}
\end{figure}

\subsection{Neutron count}

As long as the main route for the neutron absorption is the direct transition
into continuous spectrum, the neutron count at the exit is
\begin{equation}
N\left(  t\right)  =\sum N_{j}\left(  0\right)  \exp\left(  -t/\tau
_{j}^{\left(  0\right)  }\right)  . \label{c1}%
\end{equation}
The count, of course, depends on the initial number of neutrons in each state,
$N_{j}\left(  0\right)  $. When the only long-lived states are the few lowest
ones with the energies much lower than the overall kinetic energy of particles
in the beam and than the scale of the energy dispersion, it is reasonable to
assume that the initial populations of these states are the same,
$N_{j}\left(  0\right)  =N_{0}$, and
\begin{equation}
n\left(  h\right)  =N\left(  t\right)  /N_{0}=\sum\exp\left(  -t/\tau
_{j}^{\left(  0\right)  }\right)  , \label{c2}%
\end{equation}
while for each level
\begin{equation}
n_{j}\left(  h\right)  =N_{j}\left(  t\right)  /N_{j}\left(  0\right)
=\exp\left(  -t/\tau_{j}^{\left(  0\right)  }\right)  \label{cc2}%
\end{equation}
with $t/\tau_{j}^{\left(  0\right)  }$ is given by Eq. $\left(  \ref{r14}%
\right)  $. In reality, $t=L/v_{x}$ is fixed and $n$ should be considered not
a function of the time of flight $t$, but as a function of the clearance
between the walls $h$.

Below we plot the functions $n\left(  h\right)  $ and $n_{j}\left(  h\right)
$ for the overall neutron count and individual level depletions. According to
the definition $\left(  \ref{cc2}\right)  $, the neutron count from the
individual level $n_{j}\left(  h\right)  $ always changes from 1 at large $h$
to 0 at small $h$. The numerical value of the overall neutron count $n\left(
h\right)  $ also provides the information about a number of non-depleted
levels: for $n\left(  h\right)  >1$, more than 1 level is still occupied; for
$n\left(  h\right)  >2$, more than 2 levels are still occupied; for $n\left(
h\right)  >2$ - more than 3 levels are still occupied; \textit{etc.}

Since the functions $b_{j}\left(  h\right)  $ grow by orders of magnitude when
$h$ changes from 6 to 1, the ratios $t/\tau_{j}^{\left(  0\right)  }$ also
change by orders of magnitude and the neutron count $n\left(  h\right)  $
should decrease very steeply. In principle, if the absorption times $\tau
_{j}^{\left(  0\right)  }$ for different levels are vastly different, the
curve\ $n\left(  h\right)  $ $\left(  \ref{c2}\right)  $\ should become a
step-wise function with each step representing the decay of an individual
state $n_{j}\left(  b\right)  $, Eq. $\left(  \ref{cc2}\right)  $.
Unfortunately, this is not what happens.

First, we will plot the function\ $n\left(  h\right)  $ which describes the
steepest achievable depletion of the neutron population, while still remaining
within the applicability of the weak roughness theory. In other words, we will
plot $n\left(  h\right)  $ which corresponds to the largest combination of
coefficients in Eq. $\left(  \ref{r13}\right)  $. Since the weak roughness
theory is valid for roughness with the amplitude $\eta$ smaller than the
correlation radius of inhomogeneities $r$, the largest $t/\tau_{j}^{\left(
0\right)  }$ with given $r$ is achieved at $\eta=r$. The function $F\left(
\chi,\sqrt{u_{c}}r\right)  ,$ on the other hand increases with decreasing
$\chi$ though in experiment this parameter does not go below 0.15. The plot of
the function $10^{7}r^{3}F\left(  \chi,\sqrt{u_{c}}r\right)  ,$ which enters
the ratio $t/\tau_{j}^{\left(  0\right)  },$ Eq. $\left(  \ref{r13}\right)  $,
as a coefficient, is presented in Figure 9 for three values of $\chi$.
Clearly, the optimal conditions for experiment are $\chi=0.15,\ \eta=r=0.015$
when $10^{4}\eta^{2}F\left(  \chi,\sqrt{u_{c}}r\right)  \sim0.23$.%

\begin{figure}
[ptb]
\begin{center}
\includegraphics[
natheight=4.406200in,
natwidth=6.343400in,
height=3.8277in,
width=5.5002in
]%
{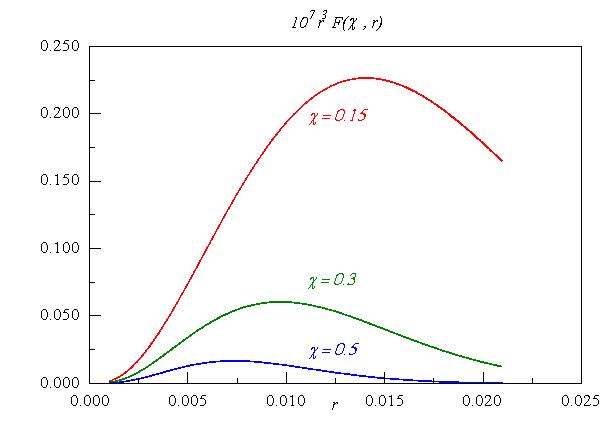}%
\caption{Function $10^{7}r^{3}F\left(  \chi,\sqrt{u_{c}}r\right)  $ for Eq.
$\left(  \ref{r14}\right)  $ at three values of $\chi,$ $\chi=0.15;0.3;0.5$.}%
\end{center}
\end{figure}

The neutron count $\left(  \ref{c2}\right)  $ with these values of $\chi
,\eta,r$ is presented in Figure 10 as a function of the interwall clearance.
The same figure also gives the individual depletions of the nine lowest lowest
quantum levels $\left(  \ref{cc2}\right)  $.%

\begin{figure}
[ptb]
\begin{center}
\includegraphics[
natheight=4.447700in,
natwidth=5.906700in,
height=4.0836in,
width=5.4137in
]%
{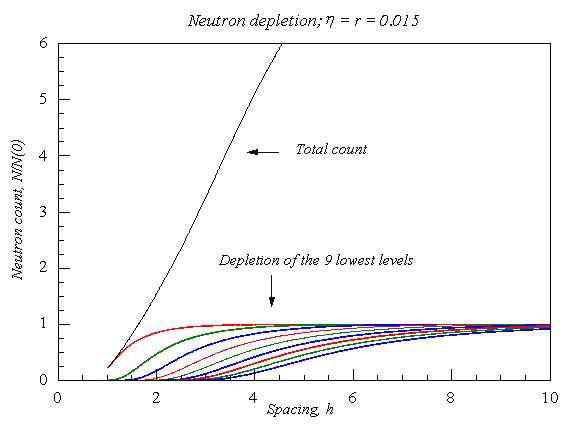}%
\caption{The neutron count $N/N\left(  0\right)  $, Eq. $\left(
\ref{c2}\right)  $, and individual depletions of the nine lowest energy
levels, Eq. $\left(  \ref{cc2}\right)  $, as a function of the spacing between
the walls $h$ at $\chi=0.15,\ \eta=r=0.01$.}%
\end{center}
\end{figure}

This figure clearly shows that though the depletion is sufficiently steep for
an unambiguous demonstration of the quantization of the gravitational states
(the total count is steep and goes to zero at $h\neq0$). However, the
"overhang" of the coefficients $b_{j}\left(  h\right)  $ [Figures 7,8] is too
large for identification of the individual levels: though the individual
depletion rates are clearly distinguishable, the overall neutron count is
still a featureless function of $h$. This "overhang" can be compensated only
by an increase in other coefficients in Eq. $\left(  \ref{r13}\right)  $. This
result means that the use of a scatterer with slight roughness is
insufficient, even under the most favorable circumstances, for an observation
of a step-wise change in neutron count and, therefore, for an identification
of higher states, unless parameter $\chi$ becomes noticeably smaller.

On the other hand, the use of large roughness such as, for example, an
increase in the amplitude of wall roughness to the value $\eta=30r$, would
modify the neutron count in Figure 10 to the one depicted in Figure 11 with
very clear steps that separate the depletions of the individual levels. Though
the parameters themselves $r=0.015,\ \eta=0.45$ are reasonable from the point
of view of experiment, the large ratio $\eta/r=30$ is out of range of the
slight roughness theory. However, the theory can be extended, as it is
explained in Section IV, to a low-amplitude high-aperture roughness and can
cover the situation plotted in Figure 11.%

\begin{figure}
[ptb]
\begin{center}
\includegraphics[
natheight=4.416600in,
natwidth=5.906700in,
height=4.1027in,
width=5.4769in
]%
{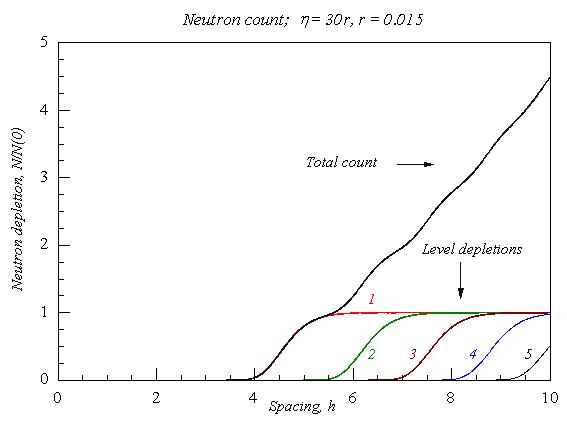}%
\caption{The neutron count $N/N\left(  0\right)  $, Eq. $\left(
\ref{c2}\right)  $, and individual depletions of the lowest energy levels, Eq.
$\left(  \ref{cc2}\right)  $, as a function of the spacing between the walls
$h$ at $\chi=0.15,\ r=0.015,\ \eta=30r$. All higher levels are already
depleted.}%
\end{center}
\end{figure}

Though the existing experiment does not produce as unambiguous picture as in
Figure 11, the similarities indicate that in experiments the roughness was not
small. Additional information on the evolution of neutron count as a function
of $\eta/r$ can be found in Figure 14 below.

\subsection{Alternative geometries}

Similar experiments have been performed in alternative geometries with, for
example, a mirror on the top and the rough scatterer on the bottom. The
theoretical description of the change in neutron population with time remains
the same as above with the only difference that all transition probabilities
are now proportional to $\left\vert \psi\left(  0\right)  \right\vert ^{2}$
instead of $\left\vert \psi\left(  H\right)  \right\vert ^{2}$. For upper
states this does not make any difference since the wave functions are
symmetric and $\left\vert \psi\left(  0\right)  \right\vert ^{2}=\left\vert
\psi\left(  H\right)  \right\vert ^{2}$. This is not so for the gravitational
states $\left(  \ref{nn14}\right)  $\ for which the wave functions on the top
and on the bottom plates can differ by orders of magnitude.%

\begin{figure}
[ptb]
\begin{center}
\includegraphics[
natheight=4.635400in,
natwidth=6.093500in,
height=4.2756in,
width=5.6126in
]%
{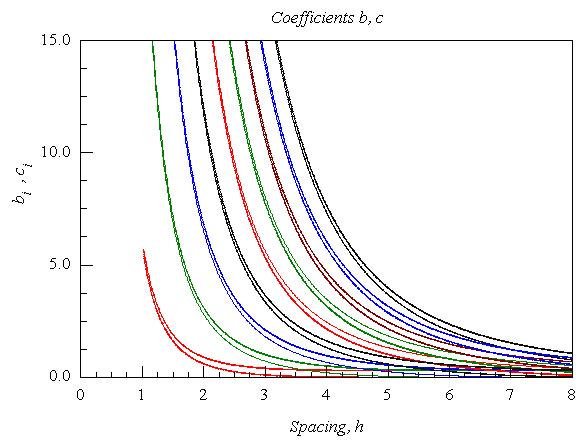}%
\caption{Pairs of coefficients $b_{j}\left(  h\right)  $ (lower curves) and
$c_{j}\left(  h\right)  $ (upper curves)$,$ Eqs. $\left(  \ref{r14}\right)  $,
$\left(  \ref{c7}\right)  $, which describe the square of the wave function on
the surfaces $s=h$ and $s=0$ for the lowest nine states. Merging of these
coefficients at small spacing between the walls $h$ correspond to the
transformation of the gravitational states into the symmetric square well
states.}%
\end{center}
\end{figure}

The ratio of the time of flight $t$ to the relaxation times $\tau_{j}^{\left(
0\right)  }$ for the gravitational states in this alternative geometry is
given, instead of Eq. $\left(  \ref{nn14}\right)  ,$ by the following
expression:
\begin{align}
\frac{t}{\tau_{j}^{\left(  0\right)  }} &  \sim10^{4}\eta^{2}\frac{(10^{3}%
r)}{10^{-3}\beta_{j}}c_{j}F\left(  \chi,\sqrt{u_{c}}r\right)  ,\label{c7}\\
c_{j} &  =b_{j}\left[  \frac{\mathrm{Ai}^{\prime}\left(  -\overline{\lambda
}\right)  -\overline{S}\mathrm{Bi}^{\prime}\left(  -\overline{\lambda}\right)
}{\mathrm{Ai}^{\prime}\left(  h-\overline{\lambda}_{j}\right)  -\overline
{S}_{j}\mathrm{Bi}^{\prime}\left(  h-\overline{\lambda}_{j}\right)  }\right]
^{2}.\nonumber
\end{align}
As expected, the coefficients $c_{j}$ are larger than the coefficients $b_{j}%
$, especially when the spacing between the walls is large. The difference
between these coefficients disappear when the spacing decreases and the states
become more symmetric as it is clearly seen in Figures 12, 13.%

\begin{figure}
[ptb]
\begin{center}
\includegraphics[
natheight=4.458100in,
natwidth=5.989700in,
height=4.1684in,
width=5.591in
]%
{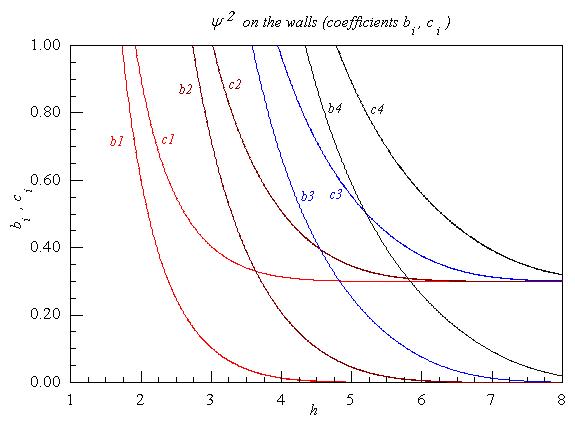}%
\caption{The same coefficients $b_{j}\left(  h\right)  $ and $c_{j}\left(
h\right)  ,$ Eqs. $\left(  \ref{r14}\right)  $, $\left(  \ref{c7}\right)  $,
for the lowest four levels as in Figure 12, but on a smaller scale. which
describe the square of the wave function on the surfaces $s=h$ and $s=0$.
Merging of these coefficients at small spacing between the walls $h$ indicate
the transformation of the gravitational states into the symmetric square well
states.}%
\end{center}
\end{figure}

When the coefficients $c_{j}$ and $b_{j}$ are comparable, the corresponding
depletion rates for both geometries are also of the same order of magnitude.
Under the same conditions as in Figure 9, though the depletion rates for the
first level in direct and inverse geometries are different because of a
noticeable difference in $c_{1}$ and $b_{1}$, the depletion rates that include
nine levels do not differ much between themselves (Figure 14; a pair of curves
marked $d-1$ and $i-1$).%

\begin{figure}
[ptb]
\begin{center}
\includegraphics[
natheight=4.076700in,
natwidth=5.280500in,
height=4.3042in,
width=5.5659in
]%
{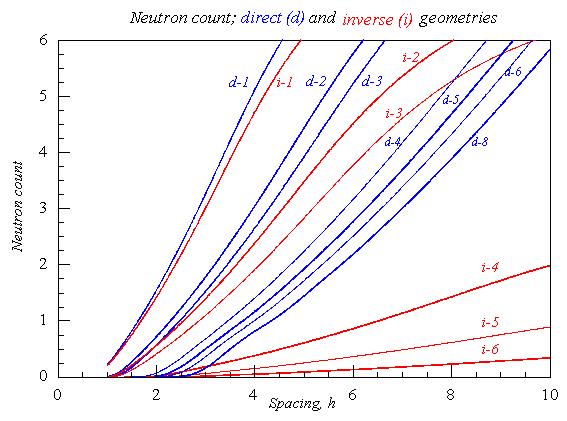}%
\caption{The depletion rates for direct ($d$; the scatterer is on the top) and
inverse ($i$; the scatterer is on the bottom) geometries for $\chi
=0.15,\ r=0.015.$ The curves are marked by the ratio $\eta/r=1;2;3;4;5;6;8$
and by the geometry, $d$ or $i$.}%
\end{center}
\end{figure}

However, with an increase in coefficient $\left(  \ref{c7}\right)  $, the
difference in depletion rates between "direct" and "inverse" geometries, which
is due to the differences between $c_{j}$ and $b_{j}$, becomes more and more
pronounced. Figure 14 contains several more pairs of curves for direct and
inverse geometries which correspond to the amplitude of roughness equal to
$\eta/r=2;3;4;5;6;8$ (the curves are marked accordingly). For even larger
roughness, such as in Figure 11, the depletion rates for the inverse geometry
is so fast (the scale is $10^{-14}$!) that the curves for direct and inverse
geometries cannot be plotted in the same figure; even for $\eta=8r$ the
numbers for inverse geometry are already too small. Note, that when the
spacing between the walls becomes small and the energy levels go
sufficiently\ high up into the square well domain, the pairs of curves, which
describe the depletion rates in direct and inverse geometries, merge between themselves.

Figure 14 illustrates the evolution of neutron count with increasing aperture
of roughness $\eta/r$. According to this figure, the increase of the ratio
$\eta/r$ from 1 to 8 is still insufficient to produce a step-wise plot of the
neutron depletion with well pronounced steps as in Figure 11. However, at
$\eta=8r$ there are already some signs of the appearance of the first step on
the depletion curve.

\section{Perspectives and conclusions}

\subsection{What to expect from strong roughness experiments?}

On the face of it, the case for shifting experiments towards strong roughness
is clear: the curves should become sharper and the chances of identifying the
individual gravitational states higher. However, as usual, there are the pitfalls.

The strong roughness can come in two flavors: large-amplitude roughness,
comparable to the spacing between the walls or to the size of the
gravitational levels, and the roughness with small lateral size of
inhomogeneities in comparison with their amplitude (the high-aperture
roughness). The use of the former one is rather dangerous. The main problem is
that the distance between the walls becomes an ill-defined variable and,
because all energy parameters are very sensitive to this spacing, the energies
extracted from such experiments are unreliable. Therefore, though the
sharpness of the curves improves, the curves, by themselves, can be used for
nothing more than qualitative identification of the individual states. The
only exception might be the so-called adiabatic roughness \cite{kaw1} in which
the profile of the walls changes so slowly that the wave functions can adjust
to the local shape of the walls. Still, even in this case the uncertainty in
energies is large making the quantitative results useless.

The low-amplitude high-aperture roughness, on the other hand, can both improve
the resolution for the lowest states and allow one to extract useful data on
the interaction between neutrons and solid surfaces. This can be achieved only
when the dominant depletion regime is the direct scattering-driven absorption
of neutrons. The moment the intermediate square-well states become involved,
the accuracy of data extracted from experiments with high-aperture walls is lost.

\subsection{Angular resolution}

Another way of increasing the sensitivity of experiment to individual levels
is to decrease parameter $\chi=U_{c}/E$ which describes the ratio of the
absorption threshold to the overall kinetic energy of particles: obviously,
the smaller is this ratio, the higher are the chances of a particle to get
absorbed after scattering. Mathematically this effect is described by a steep
function $F\left(  \chi,\sqrt{u_{c}}r\right)  ,$ Eq. $\left(  \ref{r12}%
\right)  ,$ in Figure 4.

The experimental control over the value of $\chi$ could be limited. However,
there might be an easier alternative to lowering $\chi$. In experiment, and,
as a result, in the theory above, very little attention has been paid to the
third spatial direction, $y$. What we were looking at was the
scattering-driven change of direction of particles from the beam direction $x$
to the vertical direction $z$ which could provide the particle with sufficient
kinetic energy $mv_{z}^{2}/2$ to overcome the absorption threshold $U_{c}$.
Obviously, as a result of scattering the particle acquires not only the
vertical velocity $v_{z}$, but also the component of velocity $v_{y}$ in the
third spatial direction. Though this component of velocity does not affect the
chances of the particle to be absorbed by one of the horizontal walls, the
particle can, nevertheless, escape the exit counter if the velocity $v_{y}$
and, therefore, the deviation from the initial trajectory, are large enough.
On the other hand, the exit neutron counter cannot distinguish whether the
neutrons have escaped as a result of absorption or because of angular
dispersion of the beam in $y$-direction. Therefore, instead of decreasing
parameter $\chi$, one can achieve the same effect simply by decreasing the
size of the counter in $y$-direction. Of course, the price of using the
low-angle counter will be the overall drop in the number of detected neutrons.
Therefore, following this route might require an increase in beam intensity.

From a theoretical standpoint, the use of the angular resolution requires
taking into account the scattering in $y$ direction instead of averaging in
Eq. $\left(  \ref{ee6}\right)  $. Since we have the full $2D$ expressions for
the scattering probabilities $W$, this will lead just to computational
complications rather than to conceptual difficulties. For a wide aperture
counter this extension would hardly change the nature of the results. However,
for narrower counters, as it is explained above, this might turn out to be
important. Essentially, we have to introduce a new angular counting threshold
and find the proper replacement for the function $F\left(  \chi,\sqrt{u_{c}%
}r\right)  .$ The structure of the calculations will remain the same
especially if we limit ourselves, at least in the beginning, to the
calculation of direct high angle scattering processes rather than looking at
protracted diffusion of particles over the angles.

\subsection{Contribution from the intermediate states}

In this paper we investigated only the situation when the direct scattering
into continuum dominates the neutron depletion. In principle, one should take
into account the processes that include the indirect absorption processes that
involve the intermediate energy states.

When the direct absorption processes are efficient, namely, when $r$ is
sufficiently small, adding diffusion over intermediate states should not speed
up the absorption. Involving the intermediate states can be beneficial only
for some combinations of parameters when the direct processes are suppressed
because of a relatively large value of $r$. The absorption via the
intermediate states can sometimes compensate for the presence of the
correlation exponents with $u_{c}r^{2}$ that slow down the direct absorption
processes. On one hand, the probability of scattering-driven transition
$\lambda_{i}\rightarrow\lambda_{j}$ is proportional to the square of the wave
function on the wall which decreases rapidly with lowering of the state energy
$\lambda_{j}$ and, therefore, favors transitions with as large energy increase
as possible. On the other hand, the correlation exponent $\exp\left[  -\left(
\sqrt{e-\lambda_{i}}-\sqrt{e-\lambda_{j}}\right)  ^{2}r^{2}/2\right]  $ favors
transitions into nearby states. The competition of these opposing trends
determines at what values of $r$ the intermediate states are important for absorption.

The depletion rates for higher "square well" states are always very high.
Therefore, at a relatively large value of $r$ there could even be a range of
parameters in which the depletion of the gravitational levels via the
intermediate state could become much more efficient than the direct
transitions over the threshold that the value of the threshold could become
irrelevant and the neutron count could lose its dependence on $\chi$.

In this paper we were focused on the search for "optimal" roughness that
ensures the steepest neutron depletion when the wave functions start
"touching" the walls; under these conditions, the multistage processes via
intermediate states do not seem to be necessary. However, the quantitative
description of experiment cannot be complete without a more careful look at
the role of intermediate states. We plan to address this issue later in a
separate paper.

\subsection{Short-range forces near the surface}

A fundamental issue is whether it is possible to extract the parameters of the
(ultra)short-range forces acting on neutrons near the solid surfaces (and
whether such forces exist at all). Our approach to particle diffusion along
rough walls, in which the surface roughness is treated in the same way as bulk
perturbations, is well-suited to answer the question. The short-range forces
near the surface can be considered as another $\delta-$type perturbation in
addition to the roughness-driven perturbation $\left(  \ref{nn13}\right)  $.
The matrix elements of this additional perturbation are determined by the same
equations as the ones that we used for the calculation of the surface-driven
transition probabilities. The regular part of these matrix elements determines
the shift of the eigenstates $\lambda$ and the values of the wave functions on
the walls, $b_{i}$, $c_{i}$ which are necessary for the calculation of the
neutron depletion rates. After such modification the calculation can, in
principle, proceed along exactly the same lines as it is done above. Whether
or not this program will be implemented in the future will be determined by an
experimental success in observation of a well-pronounced step-wise dependence
of the neutron count on the spacing between the walls. If this high resolution
is not achieved, the inclusion of the corrections from the short-range forces
would be meaningless.

\subsection{Conclusions}

In summary, we studied the possibility of detection of the quantized
gravitational states of neutrons. We applied our theory of quantum transport
along rough surfaces to a neutron beam between flat and rough walls and
calculated the roughness-driven transition probabilities between the quantum
states. This allowed us to found the depletion rates due to direct absorption
process as a function of the spacing between the walls and parameters of the
wall roughness. From the theoretical standpoint, the main achievement is the
extension of our theory to systems with strong, high-aperture roughness and
the application of the theory to systems with absorbing, "leaky" walls.

From the point of view of experimental applications, the focus was on finding
the optimal roughness parameters that ensure reliable resolution of several
gravitational states. The main stumbling block is the "overhang" of the wave
functions from different states which, in most situations, prevents resolution
of the nearby states. Apart from the roughness parameters, the neutron count
is also very sensitive to the ratio of the particle kinetic energy to the
absorption threshold. The most suitable experimental conditions are those in
which the characteristic times $\tau_{j}^{\left(  0\right)  }$ for the direct
transitions into continuous spectrum (absorption) are short in comparison to
the time of flight except for the lowest gravitational states for which the
probability to find the particle near the scatterer/absorber are exponentially small.

According to our results, the preferable conditions for the observation of the
well pronounced gravitational states are the following:

\begin{itemize}
\item Weak roughness is sufficient for establishing the fact of quantization,
but is not sufficient for resolving the individual quantum states. The
sharpest experimental results within the weak roughness regime are expected
when the amplitude and the correlation radius of weak roughness are $\ell\sim
R\sim0.08\ \mathrm{\mu m}$.

\item The resolution of quantum states requires the use of stronger roughness.
The use of low-amplitude high-aperture roughness $l_{0},H\gg\ell>R$ is
preferable to high amplitude roughness $\ell>l_{0}$ or $\ell>H$. In the latter
case, though the levels can be resolved, it is virtually impossible to extract
useful quantitative information from the experimental data.

\item According to our results, resolving two or three lowest states would
require $\ell\gtrsim20R$ when using low-amplitude high-aperture roughness.

\item In the case of high aperture roughness with $\eta\gg r$ one can even
neglect the correlation exponents and make the rest of the calculations as if
the roughness is weak. Physically the disappearance of these exponents means
that the changes of momenta in scattering are now unrestricted and the results
are now not very sensitive to the correlation radius.

\item The threshold velocity for the absorption should be noticeably lower
than the overall beam velocity, $\chi=U_{c}/E<1$. This condition, if necessary
could be replaced by the condition that the width of the detector in
$y$-direction $L_{y}$\ (or the width of the plates) is considerably smaller
than the distance to the detector (length of the plates) $L.$ In this case,
the absorption energy $U_{c}$ is replaced by the threshold for the
disappearance of the neutrons in $y$-direction, $E_{y}\sim\left(
L_{y}/L\right)  ^{2}E$ .
\end{itemize}

We also formulated several suggestions for future theoretical and experimental
work which could help in reliable identification of the gravitational states
and use the acquired information for the study of the short-range forces near
solid surfaces.

\begin{acknowledgement}
One of the authors (A.M.) is grateful to the ILL group for the hospitality
during his stay in Grenoble and for the stimulating discussions.
\end{acknowledgement}

\appendix

\section{Figure captions}

Figure 1. (Color online) Schematic drawing of the potential well in vertical
direction $z$. Dashed lines - the "real" potential; solid lines - the
potential used in the paper. Since $mgH/U_{c}\sim10^{-5}$ and we are
particularly interested in the lowest energy levels, the corrections are
negligible [with an accurately scaled potential, the difference between the
dashed and solid levels cannot be seen].

Figure 2. Function $F_{0}\left(  x\right)  $, Eq. $\left(  \ref{k3}\right)  $.

Figure 3. Function $F_{1}\left(  x\right)  $, Eq. $\left(  \ref{r8}\right)  $.

Figure 4. (Color online) Function $F\left(  \chi,\sqrt{u_{c}}r\right)  $, Eq.
$\left(  \ref{r12}\right)  $, at two values of $r,$ $r=0.01;0.001$.

Figure 5. (Color online) Function $F\left(  \chi,\sqrt{u_{c}}r\right)  $, Eq.
$\left(  \ref{r12}\right)  $, at three values of $\chi,$ $\chi=0.15;0.3;0.5$.

Figure 6. (Color online) Nine lowest energy levels $\lambda_{i}$ as a function
of the spacing between the walls $h$. The stars mark the positions where the
size of the energy of the level becomes equal to $mgH$ (in dimensionless
variables,\ $\lambda_{i}=h$) and the level shifts to the "square well" domain.

Figure 7. (Color online) The coefficients $b_{j}\left(  h\right)  ,$ Eq.
$\left(  \ref{r14}\right)  $, which determine the square of the wave function
on the surface of the scatterer, $B_{j}^{2}$.

Figure 8. (Color online) The same as in Figure 7 but on a smaller scale. The
stars on the curves mark the same points as in Figure 6.

Figure 9. (Color online) Function $10^{7}r^{3}F\left(  \chi,\sqrt{u_{c}%
}r\right)  $ for Eq. $\left(  \ref{r14}\right)  $ at three values of $\chi,$
$\chi=0.15;0.3;0.5$.

Figure 10. (Color online) The neutron count $N/N\left(  0\right)  $, Eq.
$\left(  \ref{c2}\right)  $, and individual depletions of the nine lowest
energy levels, Eq. $\left(  \ref{cc2}\right)  $, as a function of the spacing
between the walls $h$ at $\chi=0.15,\ \eta=r=0.01$.

Figure 11. (Color online) The neutron count $N/N\left(  0\right)  $, Eq.
$\left(  \ref{c2}\right)  $, and individual depletions of the lowest energy
levels, Eq. $\left(  \ref{cc2}\right)  $, as a function of the spacing between
the walls $h$ at $\chi=0.15,\ r=0.015,\ \eta=30r$. All higher levels are
already depleted.

Figure 12. (Color online) Pairs of coefficients $b_{j}\left(  h\right)  $
(lower curves) and $c_{j}\left(  h\right)  $ (upper curves)$,$ Eqs. $\left(
\ref{r14}\right)  $, $\left(  \ref{c7}\right)  $, which describe the square of
the wave function on the surfaces $s=h$ and $s=0$ for the lowest nine states.
Merging of these coefficients at small spacing between the walls $h$
correspond to the transformation of the gravitational states into the
symmetric square well states.

Figure 13. (Color online) The same coefficients $b_{j}\left(  h\right)  $ and
$c_{j}\left(  h\right)  ,$ Eqs. $\left(  \ref{r14}\right)  $, $\left(
\ref{c7}\right)  $, for the lowest four levels as in Figure 12, but on a
smaller scale. which describe the square of the wave function on the surfaces
$s=h$ and $s=0$. Merging of these coefficients at small spacing between the
walls $h$ indicate the transformation of the gravitational states into the
symmetric square well states.

Figure 14. (Color online) The depletion rates for direct ($d$; the scatterer
is on the top) and inverse ($i$; the scatterer is on the bottom) geometries
for $\chi=0.15,\ r=0.015.$ The curves are marked by the ratio $\eta
/r=1;2;3;4;5;6;8$ and by the geometry, $d$ or $i$.

\end{document}